\documentclass[final,authoryear,12pt,fleqn]{elsarticle}
\usepackage[utf8]{inputenc}
\usepackage{natbib}
\usepackage{amsmath,amsfonts,amssymb,amsthm}
\usepackage{graphicx}
\usepackage{rotating}
\usepackage{caption}
\usepackage{subcaption}
\usepackage{amsthm}
\usepackage{pgf}
\usepackage{setspace}
\usepackage{booktabs}
\usepackage[margin=1in]{geometry}
\usepackage{multirow}
\makeatletter
\def\ps@pprintTitle{%
 \let\@oddhead\@empty
 \let\@evenhead\@empty
 \def\@oddfoot{}%
 \let\@evenfoot\@oddfoot}
\makeatother
\usepackage{tikz,calc}
\usetikzlibrary{chains}
\usetikzlibrary{positioning}

\usepackage{accents}
\newcommand{\dbtilde}[1]{\accentset{\approx}{#1}}

\newtheorem*{nassumption}{Assumption}

\begin{document}
\begin{frontmatter}

\title{\huge{Interactive, Grouped and Non-separable Fixed Effects: A Practitioner's Guide to the New Panel Data Econometrics}\tnoteref{acknowledgements}}

\tnotetext[acknowledgements]{Acknowledgements: The authors would like to thank Aureo De Paula, Ao Wang, Anindya Banerjee, Marco Barassi, Jad Beyhum, Markus Eberhardt, Ilias Kostarakos, Giovanni Millo, Vasilis Sarafidis, Lorenzo Trapani, Chongxian Zhu and participants of the first Midlands Econometrics Conference. They would also like to thank Binzhi Chen for the grouped fixed effects estimations. All remaining errors are ours.}

\author[1]{Jan Ditzen\corref{cor1}}
\ead{jan.ditzen@unibz.it}
\author[2,3]{Yiannis Karavias} \ead{yiannis.karavias@brunel.ac.uk}
\cortext[cor1]{Corresponding author}
\affiliation[1]{Free University of Bozen-Bolzano}
\affiliation[2]{Brunel University of London}
\affiliation[3]{University of Birmingham}

\begin{abstract}
The past 20 years have brought fundamental advances in modeling unobserved heterogeneity in panel data. Interactive Fixed Effects (IFE) proved to be a foundational framework, generalizing the standard one-way and two-way fixed effects models by allowing the unit-specific unobserved heterogeneity to be interacted with unobserved time-varying common factors, allowing for more general forms of omitted variables. The IFE framework laid the theoretical foundations for other forms of heterogeneity, such as grouped fixed effects (GFE) and non-separable two-way fixed effects (NSTW). The existence of IFE, GFE, or NSTW has significant implications for identification, estimation, and inference, leading to the development of many new estimators for panel data models. This paper provides an accessible review of the new estimation methods and their associated diagnostic tests and offers a guide to empirical practice. In two separate empirical investigations we demonstrate that there is empirical support for the new forms of fixed effects and that the results can differ significantly from those obtained using traditional fixed effects estimators.\\
\end{abstract}

\end{frontmatter}

\clearpage
\doublespacing
\section{Introduction}
\label{section::intro}
In the almost 60 years that came after the seminal paper by \cite{BalestraNerlove1966}, the merits of panel data sets in dealing with endogeneity due to omitted variables have been well understood, and this has led to the standard practice of consistent repeat observation and measurement of many units over long periods. The availability of large $N$ - large $T$ panel datasets, where $N$ denotes the number of units and $T$ denotes the number of time series observations, is now pervasive in all strands of the economics literature and beyond.\footnote{See \cite{Baltagi2008} and \cite{Hsiao2014} for textbook introductions to the panel data literature.}

A fundamental assumption in panel data models is the inclusion of individual effects to capture the effect of time-invariant unobserved characteristics, which are effectively unobserved omitted variables. Observation over long periods, however, possibly over years and decades, puts a strain on the sufficiency of the time invariance assumption, as the omitted variables are more likely to have time-varying effects instead. This argument is conceptually no different from that which led to the emergence of the structural breaks literature, where the long time series dimension increased the probability of time-varying parameters as a result of changes in the economic conditions \citep{Perron1989,Andrews1993}. As such, models and methods for capturing time-varying unobserved heterogeneity became necessary. The two-way fixed effects (TWFE) model, which is the current workhorse in the literature, partly addresses this issue through a time-varying common effect. Still, this is not sufficient, as it imposes the assumption that time variation affects equally all units, which is hard to justify given unit heterogeneity.

A key modeling innovation, developed for different reasons but nevertheless capable of dealing with time variation, is the introduction of ``interactive fixed effects'' (IFE) by \cite{Holtz-Eakin1988,RobertsonSymons2000,AHN2001219}, and \cite{coakleyfuertessmith2002}. The IFE represents a significant novelty and generalization compared to the TWFE, in that it allows individual effects to have a time-varying and heterogeneous impact on the dependent variable. It is useful to compare TWFE and IFE. Starting with a TWFE model for the errors $ u_{i,t}=z_i+f_t+\varepsilon_{i,t} $, let $ z_i $ be the individual effect and $ f_t $ be the factor or time effect. The index $i=1,...,N$ where $N$ is the number of units in the panel and $t=1,...,T$ where $T$ is the number of time series observations. A simple version of the IFE model then is $ u_{i,t}=z_if_t+\varepsilon_{i,t} $. The interaction term $z_if_t$ is the source of the ``interactive'' fixed effects nomenclature, which is due to \cite{Bai2009}. The IFE allows the $ z_i $ and $ f_t $ to be $ m $-dimensional vectors, in which case TWFE can be a part of IFE for $z_if_t=z_{i,1}+f_{1,t}+z_{i,2}f_{2,t}+...+z_{i,m}f_{m,t}$, or even the only part if $m=2$.

The main motivation behind this paper is to introduce the new estimation methods and to nudge applied panel data research towards the greater use of the IFE and its extensions, given that to this date, it is somewhat limited. Despite IFE being a key topic in theoretical research, its use is dysanalogous in empirical research. One reason for this could be the lack of understanding of the pervasiveness of this type of endogeneity when $T$ is large, while another one could be the only recent development of many methods that are used to estimate IFE models. In this paper, we review the key contributions in the area in terms of estimation and diagnostic testing and discuss the basic assumptions and their implications for empirical practice. Our focus lies on estimation methodology and the assumptions that have a direct impact on the choices of the researcher, and not on technical regularity conditions, which will be omitted. To further facilitate the use of IFE models, an appendix with a list of references to readily available software is included at the end. As such, this paper complements previous literature reviews such as \cite{Hsiao2014,ChudikPesaran2015}, \cite{JuodisReese2025} and \cite{bonhommedenis2025}. 

The introduction of IFE is followed by two more new forms of unobserved heterogeneity: group fixed effects (GFE) \citep{Bonhomme2015a} and non-separable two-way heterogeneity (NSTW) \citep{bonhommelamadonmanresa2022}. The GFE model is a special case of IFE where the $z_i$ are indicator functions such that each of the $m$ factors is unique for each of the $m$ distinct groups in the data. GFE can be preferable to IFE in smaller $T$ datasets, as it is a more parsimonious model restricting the $z_i$ from varying across the $N$ units to varying only across the $m$ groups. NSTW is a, potentially nonlinear, generalization of IFE and can be written as $u_{i,t}=h(z_i,f_t)+\varepsilon_{i,t}$, where $h(\cdot,\cdot)$ is a smooth function, unknown to the researcher. It can be shown that $h(z_i,f_t)$ can be approximated by an IFE model with $m\to\infty$ \citep{FREEMAN2023105498}. The emergence of GFE and NSTW marks an important shift in estimation methodology and also in the key econometric difficulties faced, which are different from those that have been accompanying TWFE. NSTW models are estimated by a new method, termed discretization, and introduced in \cite{bonhommelamadonmanresa2022}. Discretization was inspired by GFE and involves approximating the unknown function $h(z_,f_t)$ with clustering. Once the clusters have been estimated, then the two-way clustered fixed effects estimator can be used. Overall, including IFE, the key issues in the new estimation methods include i) dealing with the unknown nuisance parameter $m$, ii) the minimization of appropriate convex objective functions instead of the sum of squared residuals subject to a constraint, and iii) clustering methods and computational difficulty. 

In a set of two empirical applications, we apply a host of the new estimation methodologies and diagnostics that we present, demonstrating their applicability and usefulness. The IFE, GFE, and NSTW estimators are compared to the TWFE estimator. The key finding is that TWFE is not sufficient for modeling the unobserved heterogeneity, and potentially neither is IFE. Instead, the diagnostic tests in a barrage of different regressions in both applications confirm that NSTW is most appropriate, hence evidencing, for the first time, the existence of nonlinear forms of unobserved heterogeneity.  

Our aim is to present the new estimation methodologies in an intuitive way. To simplify the exposition, we consider estimation only in the simple linear regression model. As such, we will not present extensions to the estimation methods applied in areas such as limited dependent variable models and models for program evaluation, multivariate panel models, spatial panels, heterogeneous/time-varying coefficient models, non-stationary panels, forecasting, and high-dimensional panels.

The paper is organized as follows. Section 2 offers a historical empirical motivation for IFE, aiming to demonstrate that their existence is justified by economic theory—and that attempts were made to estimate models with IFE before the current econometric apparatus became available. Section 3 presents the main estimators in the literature and discusses their most important assumptions. Section 4 presents diagnostic tests and tests of model specification. Section 5 includes the two empirical applications, while Section 6 concludes the paper. The appendix includes a catalog of readily available software in Stata and R.

\section{Empirical Motivation of IFE}
Having in mind the applied researcher, this section argues that IFE is widely useful in panel data research, whether in microeconometrics, macroeconometrics, financial econometrics, or beyond. IFE is not simply a deductive theoretical extension void of empirical content, but it has a long history in economic thought and applied research. Of course, the early attempts had significant restrictive assumptions. IFE is a linear combination of prices or factors, and this idea can be traced back to classical consumer theory; see, e.g., \cite{lancaster1966}. In empirical research, \cite{hause1980} argued that the large standard errors observed in earnings regressions indicated significant unexplained variability, explaining the term ``unexplained'' in that earnings, and thus worker productivity, could not be captured by observed individual characteristics. He thus proposed an interactive effects model with the unobserved individual characteristics ($z_i$) being ``economic ability'' and ``on-the-job training''. While economic ability was considered time-invariant, on-the-job training was interacted with $f_t=t$, capturing the additional skills that workers eventually get access to, given their on-the-job training during their early years in the labor market. \cite{heckmanscheinkman1987} suggest that earnings can be decomposed into payments to separate worker productive characteristics. In our notation, they assume that $z_i$ is a vector of observable worker characteristics and $f_t$ is the vector of prices of these characteristics across sectors, where $t$ here denotes a specific sector. They proceed to reject the hypothesis that these characteristics are uniformly priced across sectors with $f_t=f$ for every $t$, empirically rejecting the standard time-invariant fixed effects model. 

In the micro-econometric setting of returns to schooling, $ z_i $ is typically the unobserved individual effect capturing time-invariant characteristics like innate ability, conscientiousness, conformity, self-esteem,  or other soft skills. However, the price of these skills can vary over long periods and this can be modeled by interacting the vector $z_i$ with the vector of time-varying prices $ f_t $. Both skills and their prices are allowed to be correlated to the main variable of interest, the years of education; see e.g., \cite{kejriwal2020}. Finally, on the adjacent topic of the impact of minimum wage on employment or wages, $ z_i $ captures permanent differences across states and $ f_t $ captures aggregate labor market conditions. Interactive fixed effects allow for the possibility that different states have different aggregate labor market conditions every year. \citet[page 7]{manning2021} presents seven regression models with various fixed effects formulations and trends ``all intended to capture the possibility that the evolution of labor market conditions may vary across states in a complicated way''. All the above examples demonstrate the need for a richer unobserved heterogeneity structure, which is ably captured by the richness of the full $ m $-dimensional IFE model $ u_{i,t}=z_{i,1}f_{1,t}+...+z_{i,m}f_{m,t}\varepsilon_{i,t} $.

In order to demonstrate the pervasiveness and necessity of IFE, a few more examples are in order. \cite{RobertsonSymons2000}, \cite{coakleyfuertessmith2002} and \cite{Pesaran2006} employed the IFE model as a means of capturing cross-section dependence. This is particularly relevant in a growth regression framework, where $ z_i $ can capture permanent differences between countries, like geographical characteristics, and $ f_t $ can capture the varying price of these characteristics as technology and political frontiers evolve. Alternatively, $ f_t $ can be business cycle fluctuations, innovations in technology, productivity shocks, risk premia, unobserved price changes or other general shocks such as global market volatility, while $ z_i $ is the heterogeneous country response arising from geographical, institutional and cultural idiosyncrasies, see e.g. \cite{ChudikPesaran2015} and \cite{eberhardtteal2013}. In financial economics, \cite{miranda2020} consider US monetary policy as a global factor in asset prices. \cite{cesa2020} assume that each county's GDP growth is affected by a common factor which captures world growth driven by productivity. \cite{mumtazsurico2009} study the transmission of international shocks into the UK using a factor augmented VAR model and argue that international factors are generated from real activity, inflation, liquidity, and comovements in short term interest rates.

Clearly, the IFE model unifies much of the existing panel data literature under a common framework and allows for even more robust results. For those that still remain skeptical, unwarranted inclusion of IFE, or the even more general NSTW, causes only efficiency loss which may not be a serious issue in applications with high data availability; the alternative of faulty exclusion of IFE causes inconsistency \citep{kimoka2014}. Based on the above, if one accepts the need for the IFE, then the same argument of unit-specific time variation of unobserved heterogeneity also motivates the GFE and NSTW models, each with its own specific strengths. GFE is a parsimonious model that yields efficiency gains in the presence of a group structure, while NSTW generalizes IFE in nonlinear settings or in settings where $m\to \infty$. Group structures and nonlinearity in economic data are well established in the literature, so we refrain from further elaboration.   

\section{New Forms of Unobserved Heterogeneity}
\subsection{Interactive Fixed Effects}
We introduce interactive fixed effects as a departure from the well-known individual effects formulation. Consider the linear panel data model:
\begin{align}
y_{i,t} &= x'_{i,t} \beta + c_i+ \varepsilon_{i,t} \label{eq:main}
\end{align}
where $x_{i,t}$ is a $K \times 1$ vector of observed covariates and $\beta$ is a $K \times 1$ vector containing the main parameters of interest.\footnote{In the following we assume that $K$ is fixed, therefore, we will not consider high-dimensional panel data, see e.g. \cite{linton2022}.} The scalar $c_i$ is the individual effect and includes the constant regressor and other time-invariant variables $z_{i,j}$:
\begin{align}\label{eq:fe2}
c_i=z_{i,1}f_1+...+z_{i,m}f_m=z_i'f.
\end{align}
The vector $z_i$ includes time-invariant observed variables such as the constant, sex, ethnicity, and geographic location, but also variables that are usually unobserved, such as motivation, innate ability, technology, managerial skills, geography, productivity, and more. Because some of these variables are unobserved, the whole $z_i$ is considered as unobserved.

The $m\times 1$ vector $f$ contains slope coefficients of the variables in $z_i$. A stringent assumption in \eqref{eq:fe2} is that the $ f $, are constant in time. The IFE model is obtained by allowing $f$ to be time-varying:
\begin{align}\label{eq:ie}
c_{i,t}=z_{i,1}f_{1,t}+...+z_{i,m}f_{m,t}=z_i'f_t.
\end{align}
Some parallelisms between fixed effects (FE) and IFE are in order.\footnote{In the text we follow \cite{mundlak1978} and \cite{wooldridge2010} and define the term ``fixed'' to mean that the unobserved effects are allowed to be correlated with the regressors.} FE can be correlated with the regressors $x_{i,t}$, and this is still the case with IFE. FE are rarely the main parameters of interest in empirical research and this also applies to IFE. Finally, like FE, IFE must not be ignored because this can lead to biased and inconsistent estimators.

A nuisance parameter that will shape much of the discussion below is $m$, the number of factors. In most applications $m$ is fixed and unknown-yet a fixed $m$ imposes a restriction on the number of individual effects with time-varying prices. This restriction is relaxed by NSTW, discussed further below.

Before we proceed to estimation, it is useful to rewrite \eqref{eq:main} and \eqref{eq:ie} in matrix notation, where we stack across time:
\begin{align}\label{mod:ifemat}
y_i&=X_i\beta+u_i\\
u_i&=Fz_i+\varepsilon_i,\label{mod:ifemat2}
\end{align}
where $y_i=(y_{i,1},y_{i,2},...,y_{i,T})'$, $u_i=(u_{i,1},u_{i,2},...,u_{i,T})'$, and $\varepsilon_i=(\varepsilon_{i,1},\varepsilon_{i,2},...,\varepsilon_{i,T})'$ are $T\times 1$ vectors, $\beta$ is a $K\times 1$ vector of coefficients, $F=(f_1,f_2,...,f_m)'$ is a $T\times m$ matrix of unobserved common factors and $z_i=(z_{i,1},...,z_{i,m})'$ is a $m\times 1$ vector of unobserved factor loadings. In the following, we will assume that $F$, $z_i$, and $X_i$ are independent of $\varepsilon_i$, while $Fz_i$ and can be dependent to $X_i$.\footnote{If $X_i$ is correlated to $\varepsilon_i$ then we have an additional source of endogeneity, beyond that of IFE. This case requires external instruments to be available; see \cite{baltagifengkao2019} and \cite{hongsujiang2023}.} 

If $T>N$ the system can be treated as Seemingly Unrelated Regressions and be estimated using the feasible generalized least squares estimator of \cite{zellner1962} and the correlation across units can be even more general than the IFE model. However, if $N>T$, as is usually the case in panel data, this approach becomes infeasible. If, however, the factors in $F$ were observed, then the consistent ordinary least squares (OLS) estimator would be
\begin{equation}\label{est:ools}
\hat \beta_{IFE}=\left(\sum_{i=1}^NX_i'M_FX_i\right)^{-1}\left(\sum_{i=1}^N X_i'M_F y_i\right),
\end{equation} 
where $M_F$ is the orthogonal projection matrix $M_F=I_T-F(F'F)^{-1}F'$ and $I_T$ is the $T$-dimensional square identity matrix. This becomes the within-groups or fixed effects estimator when $ F=(1,1,...,1)' $.  Unfortunately, the factors in $F$ are usually not observed and the above OLS estimator is infeasible.

\subsection{Grouped Fixed Effects}
Efficiency gains can be obtained if the unobserved heterogeneity does not vary across units but rather across groups of units. Grouped fixed effects are a special case of interactive fixed effects and are reasonable to assume in settings where clusters naturally arise, i.e. in classrooms, schools, or countries; see, e.g., \cite{mackinnon2023cluster}. Economic theory also predicts the existence of latent group structures, for example, in the form of multiple equilibria, which are the outcomes of game-theoretic models \cite{BajariHahnHongRidder2011}. 

We follow here the GFE literature and assume that there are $ G $ groups in the population, and unit $ i' $s membership is given by $ g_i=g $ where $ g\in \{1,2,...,G\} $. For example, if unit $ s $ belongs to group $ j $, then $ g_s=j $. In this case, \eqref{eq:ie} becomes:
\begin{align}\label{eq:gfe}
	c_{i,t}=I\{g_i=1\}f_{1,t}+...+I\{g_i=G\}f_{G,t},
\end{align}
where $ I\{\cdot\} $ denotes the indicator function. The GFE are a special case of IFE with factor loadings equal to the group membership indicator function and each factor loading only a specific group of units. In the above expression this can be seen by setting $ z_{i,g}=I\{g_i=g\} $ and  $ G=m $. 

The GFE specification has nontrivial implications for model estimation as the number of individual effects is no longer of order $ O(N) $ but of $ O(G) $ where $ G $ is a small and fixed number. Therefore it avoids the incidental parameters problem or the Nickell bias in dynamic panels. If group membership is known, then the GFE estimator of $\beta$ is given by \eqref{est:ools} with the appropriate simplification due to the group membership.

\subsection{Non-separable Two-Way Heterogeneity}
\cite{bonhommelamadonmanresa2022} consider a general model of unobserved heterogeneity: 
\begin{equation}\label{eq:nstw}
    c_{i,t}=h(z_i,f_t),
\end{equation}
where $h(\cdot,\cdot)$ is a smooth real valued function. For $h(\cdot,\cdot)=z_i'f_t$ the above model nests IFE as a spcial case, but $h(\cdot,\cdot)$ can be non-linear, and as such NSTW is more general over IFE. For example, the unobserved heterogeneity can now take a constant relative risk aversion, or a constant elasticity of substitution form, i.e. $h(z_i,f_t)=(d z_i^\gamma+(1-d) f_t^\gamma)^{1/\gamma}$, for $m=1$. 
 
Furthermore, NSTW addresses the restriction in IFE that the number of individual effects with changing prices $m$ is not allowed to be large compared to $N$ and $T$, and which can in turn limit the applicability of the model. With traditional fixed effects a single $z_i$ can capture an infinite number of time invariant unobserved characteristics, therefore the number of individual effects is not restricted. If, however, the prices of these characteristics change in a heterogeneous manner, then the number of such characteristics in the model must be limited, becoming a restriction. Consider for example the individual effects in a growth regression; these include law system and judicial efficiency, colonial past and history of institutions, political systems, cultural norms, landlocked status, ethnic fractionalization, linguistic differences, climate zone and natural resource endowment, to name a few. Climate change, advancements in technology and changing geopolitical alliances can increase the price of some characteristics and reduce it for others, therefore, potentially making the IFE model with a fixed $m$ inappropriate. The longer the period of observation, the higher the probability that many individual characteristics experience a change in price variation. An early contribution by \cite{ChudikPesaranTosetti2011} considered the model:
\begin{equation}\label{eq:ime}
    c_{i,t}=\sum_{j=1}^m z_{i,j}f_{j,t}+\sum_{j=m+1}^{N} z^{w}_{i,j}f^{w}_{j,t}, 
\end{equation}
where there are two types of factors, the $m$ strong ones $f_{j,t}$ and the $N-m$ weak ones $f_{j,t}^w$, which become infinite as $N\to \infty$.\footnote{A factor $f_{j,t}$, for $j=1,...,m$, is strong if $lim_{N\to\infty}\sum_{i=1}^N|z_{i,j}|=C>0$ and weak if $lim_{N\to\infty}\sum_{i=1}^N|z_{i,j}|=C<\infty$ \citep{ChudikPesaranTosetti2011}. Strong factors affect all units in the panel while weak factors only some. In other words, an individual effect $z_{i,j}$ has a time varying price $f_{j,t}$ for some units and not for all, if $f_{j,t}$ is a weak factor.} It was further assumed that the weak factors could not be correlated with the regressors. This would mean that they are ``interactive random effects'', and in total we would name model \eqref{eq:ime} as ``interactive mixed effects'', being a combination of IFE and interactive random effects. Having the infinite weak factors is not restrictive in practice \citep{Onatski2010,FREEMAN2023105498}, but assuming them uncorrelated to the regressors is.

Under weak regularity conditions, however, \eqref{eq:nstw} is equal to an infinitely large IFE model:
\begin{equation}
    c_{i,t}=\sum_{j=1}^\infty z_{i,j}f_{j,t}.
\end{equation}
without any restrictions in the correlation of $c_{i,t}$ and the regressors \citep{FREEMAN2023105498}.
Therefore, the non-separable unobserved heterogeneity model can be seen as an IFE model with infinite individual effects and factors.\footnote{\cite{GaoLi2024} also consider infinite-dimensional fixed effects but in a panel multinomial choice model.} 

\section{Next-Generation Panel Data Estimators} \label{sec:estim}
\subsection{Estimators for Panel Data with Interactive Fixed Effects}\label{sect:IFE}
In the late 1990s, cross-sectional dependence emerged as a prominent concern in panel data unit root tests, used in studies of growth convergence and the purchasing power parity hypothesis \citep{OConnell1998}. The IFE structure inherently implies unit cross-sectional dependence, and for this reason, it was employed in \cite{coakleyfuertessmith2002} and \cite{Pesaran2006} in general panel data regressions. The presentation below does not follow the historical development of the estimators; instead, it is organized thematically around the assumptions required for each one. 

\subsubsection{The Iterative Least Squares / Quasi Maximum Likelihood Estimator}
\cite{Bai2009} and \cite{moonweidner2017} suggest treating $z_i, \; i=1,...,N$ and $F$ as unknown parameters that can be estimated by Iterative Least Squares (ILS) at the same time as  $\beta$.\footnote{Other names for the ILS estimator in the literature include ``principal components'' or ``least squares estimator'' or the ``concentrated least squares estimator''. \cite{moonweidner2017} show that the ILS estimator minimizes the Gaussian quasi maximum likelihood function and therefore use the name quasi maximum likelihood estimator. A final name is the ``profile least squares estimator'', mentioned in \cite{moonweidner2019}, where ``profile'' is another word for ``concentrated''.} The iteration is based on the insight that if $ F $ is known, then $ \beta $ can be consistently estimated by the oracle estimator in \eqref{est:ools}. Equally, if $ \beta $ is known, then $ F $ can be estimated by principal components (PC) in the factor model  $y_{i}^\star=Fz_i+\varepsilon_i$, where $y_{i}^\star=(y_i-X_i\beta)$. Therefore, the ILS estimator is the iterative solution to the set of non-linear equations:
\begin{align}\label{eq:PCbetaestimation}
\hat \beta_{ILS}=\left(\sum_{i=1}^NX_i'M_{\hat F_{ILS}}X_i\right)^{-1}\left(\sum_{i=1}^N X_i'M_{\hat F_{ILS}} y_i\right),
\end{align}
\begin{align}\label{eq:PCFestimation}
	\left[\frac{1}{NT}\sum_{i=1}^N \left(y_i-X_i\hat\beta_{ILS}\right)\left(y_i-X_i\hat\beta_{ILS}\right)'   \right]\hat F_{ILS}=\hat F_{ILS} \hat V_{NT},
\end{align}
which involves iterating estimation of $ \beta $ and $ F $ until $ \hat \beta_{ILS} $ and $ \hat F_{ILS} $ converge.  $ \hat V_{NT} $ is the diagonal matrix consisting of the $ m $ largest eigenvalues, arranged in decreasing order, of the matrix $ \left[(NT)^{-1}\sum_{i=1}^N (y_i-X_i\hat\beta_{ILS})(y_i-X_i\hat\beta_{ILS})'\right] $. Once $ (\hat\beta_{ILS}, \hat F_{ILS}) $ have been obtained, it is further possible to estimate the factor loadings $ z_i $ according to $ \hat z_{i,ILS}=(\hat F_{ILS}'\hat F_{ILS})^{-1}\hat F_{ILS}'(y_i-X_i\hat\beta_{ILS}) $ for all $ i $. Two influential features in this process are the initial estimate of $\beta$ which begins the iteration process, and the number of factors $m$. Both choices will be discussed further below.

The ``least squares'' characterization is derived from the fact that $ (\hat\beta_{ILS}, \hat F_{ILS}, \{\hat z_{i,ILS}\}_{i=1}^N) $ minimizes the sum of squared residuals function:
\begin{equation}\label{eq:ssr}
	SSR(\beta, F, \{ z_{i}\}_{i=1}^N)=\sum_{i=1}^N(y_i-X_i\beta-Fz_i)'(y_i-X_i\beta-Fz_i).
\end{equation}
Unfortunately, there is no closed-form solution for $\hat\beta_{ILS}$ and the number of iterations until convergence can vary, and with it, computational time.\footnote{The alternative estimation methodology of taking first-order conditions in \eqref{eq:ssr} and solving for the unknowns, instead of the iteration, is computationally more tedious.}  

The ILS estimator requires only weak assumptions and it places no restriction on the relationship between IFE and the regressors. If $N,\; T\to \infty$ with $ T/N\to \rho>0 $, and the following assumptions hold: 
\begin{nassumption}[ILS Estimator Factor Requirement]\label{ass:ls_f_m}
	$m$ is known to the researcher. 
\end{nassumption}
\begin{nassumption}[Strong Factors]\label{ass:ls_f_f}
\begin{align}
	    \underset{T\to\infty}{lim}\frac{1}{T}\sum_{t=1}^T f_tf_t'&=\Sigma_F>0\; \text{for some}\; m\times m \; \text{matrix} \; \Sigma_F,\\
     \underset{N\to\infty}{lim}\frac{1}{N}\sum_{t=1}^T z_iz_i'&=\Sigma_Z>0\; \text{for some}\; m\times m \; \text{matrix} \; \Sigma_Z.
\end{align} 
\end{nassumption}
\noindent Then, it shown that the $ILS$ estimator achieves the parametric rate of convergence $NT$, but is asymptotically biased \citep{Bai2009,moonweidner2017}:
\begin{equation}
	\sqrt{NT}(\hat\beta_{ILS}-\beta)\xrightarrow{d}N(B, V),
\end{equation}
where the bias $B=-\rho^{-1/2}B_1-\rho^{1/2}B_2-\rho^{-1/2}B_3$. The first of the bias terms $B_1$ arises only in the presence of lagged regressors and is a more general form of the Nickell bias. The second term $B_2$ arises only if there is cross-sectional heteroskedasticity and/or weak cross-correlation in $\varepsilon_{i,t}$ across $i$. $B_3$ arises only when there is time-series heteroskedasticity and/or serial correlation in $\varepsilon_{i,t}$.\footnote{Notice that the errors cannot be serially correlated in the presence of lagged dependent variables.} If the regressors are all exogenous and the errors $ \varepsilon_{i,t} $ are i.i.d. over time and across units, then $ B_1=B_2=B_3=0 $ and $ \hat\beta_{ILS} $ is unbiased. \cite{Bai2009} and \cite{moonweidner2017} suggested a modified, unbiased, $ILS$ estimator to correct the asymptotic bias, the estimator is given by 
\begin{equation}\label{eq:ilsbc}
    \hat\beta_{ILS,Bc}=\hat \beta_{ILS}+T^{-1}\hat B_1+N^{-1}\hat B_2+T^{-1}\hat B_3,
\end{equation} 
which is still $\sqrt{NT}$-consistent. 

A restrictive assumption in the application of the ILS estimator is that the number of factors $m$, is known. Ideally $m$ should be estimated at the same time as $\beta$, but this is a difficult problem, and for this reason theoretical results in the literature are based on the assumption that $m$ is known. This is not sufficient in practice, and consistent estimation of $m$ is possible, but potentially unreliable, as can be seen in the Monte Carlo section of \cite{Onatski2010} and in the empirical application of \cite{moonweidner2015}, in which the number of estimated factors ranges from 1-9. The estimation of $m$ is further impacted by the initial estimate for $\beta$ in the first step of the iteration. Both these estimators impact the small sample behavior  of the $ILS$ estimator \citep{westerlundurbain2015}. 

Under the assumption of strong factors, which is used by the PC estimator of $F$, estimating $m$ can be avoided, as long as a $\hat m$ is available, such that $\hat m\geq m$.\footnote{It is common practice in applied econometrics to select a parameter so that it is greater than the true one. For example, selecting the lag order of VAR or the lag length in a covariance matrix estimator \citep{neweywest1987}. Selecting the parameter larger than the true one leads to inefficiency but not to inconsistency - the same applies here for $\hat m<m$.} \cite{moonweidner2015} show that the ILS estimator based on the arbitrarily selected $\hat m$ has the same distribution as the ILS estimator based on $m$. In other words, there is no cost in efficiency from including irrelevant factors in the estimation. If however, the factors are weak, then the ILS estimator has a slower, $\sqrt{min(N,T)}$, rate of convergence, see also \cite{baing2023}.

The assumption of strong factors implies that the factors in \eqref{eq:ie} are pervasive across units. This is also a statement on the strength of the factor loadings across units. In microeconomic studies frequently all factors are strong. For example, the productivity of a person depends on innate ability, and innate ability is ``important'' for every person. In financial and macroeconometric studies, however, it is possible that some factors load only to some of the units. Examples of this appear in asset pricing, see e.g. \cite{chamberlainrothchild1983}, \cite{connorkorajczyk2024} and \cite{anatolyevmikusheva2022}, and in capturing cross-section dependence, see e.g. \cite{ChudikPesaranTosetti2011}. \cite{ArmstrongWeidnerZeleneev2025} show that in the presence of weak factors, ILS becomes biased and non-normally distributed because weak factors cannot be distinguished from noise. They propose a debiased estimator with confidence intervals which remain valid for any factor strength. 

In terms of the number of iterations, \cite{jiang2021} argue that, under some regularity conditions, the iterative estimator becomes consistent after only one iteration. However, it is possible that the iteration scheme does not converge, as in \cite{castagnettirossitrapani2019}, which employ a panel with $N=207$ and $T=33$. \cite{jiang2021} note that iterations can have diminishing returns and so they suggest an optimal number of iterations which can be used as a stopping rule. 

The need for an initial estimator and an estimator of $m$, and the number of iterations can lead to poor small sample performance of the ILS estimator, an issue documented in \cite{westerlundurbain2015} and \cite{Juodis2022a}. Its performance deteriorates especially for dynamic panel data models. A suggestion is that results should be presented for various values of $ \hat m $ as a robustness check, in order to gauge the impact of additional factors on the estimates. If $ \hat m>m $, it is expected that a greater $ m $ will not be accompanied with significant changes to the coefficients, however, if $ \hat m<m $, increasing $ \hat m $ should change the estimates considerably, as unobserved heterogeneity is removed. A final caution is that the $SSR$ in \label{eq:ssr} is not convex under $ m<\hat m $, an issue which is dealt with by the Post Nuclear Norm Regularized estimator presented below.

\subsubsection{The Penalized Least Squares Estimator}
Another solution to the problem of unknown $m$, is to estimate the parameter at the same time as $\beta$. \cite{LuSu2016} propose the Penalized Least Squares (PLS) estimator, which estimates the number of factors using adaptive group LASSO. Under the assumption of strong factors, the PLS estimator has the oracle property, meaning that it is asymptotically equivalent to ILS with $m$ known. The PLS estimator is calculated as follows:\footnote{The exposition in \cite{LuSu2016} includes a second penalty term, which allows the selection of relevant regressors out of a large number of available ones. We omit this term for ease of exposition and also because we do not consider the scenario $K\to \infty$ here.}
\begin{enumerate}
    \item Step 1: ILS. Select a large $\hat m$ such that $m\leq \hat m$, and obtain $ (\hat\beta_{ILS, Bc}, \hat F_{ILS,Bc}, \{\hat z_{i,ILS,Bc}\}_{i=1}^N) $ and the residuals $\hat u_{i}$, where $\hat F_{ILS,Bc}$ and $\{\hat z_{i,ILS,Bc}\}_{i=1}^N$ are the final estimates based on $\hat\beta_{ILS, Bc}$. The $\hat m$ estimated factors $\hat F_{ILS,Bc}$ include $m$ true factors and $\hat m-m$ redundant ones.  
    \item Step 2: Eigenvalue estimation. Estimate the matrix $\hat\Sigma_{\hat F}=T^{-1}\hat F'\hat F$, where $\hat F=(NT)^{-1}\sum_{i=1}^N \hat u_i \hat u_i'\hat F_{ILS,Bc}$. Obtain the $\hat m$ eigenvalues $\lambda_1,...,\lambda_{\hat m}$ of $\hat\Sigma_{\hat F}$, arranged in descending order. The $\hat m - m$ smaller ones should asymptotically converge to $0$. 
    \item Step 3: Penalization. Minimize the penalized sum of squared residuals:
    \begin{equation}
        Q_{\phi}(\beta,\{z_{i}\}_{i=1}^N)=SSR(\beta, \hat F_{ILS,Bc}, \{ z_{i}\}_{i=1}^N)+\frac{\phi_{NT}}{\sqrt{N}}\sum_{j=1}^{\hat m}\frac{1}{\lambda_j^2}\|z_{N,j} \|,
    \end{equation}
    where $z_{N,j}$ is the $j$-th column of the $N\times \hat m$ matrix $z_N=(z_1',z_2',...,z_N')'$ of factor loadings, $\|\cdot \|$ is the Frobenius norm and $\phi_{NT}$ is a tuning parameter chosen by an information criterion.
\end{enumerate}
The final $\hat\beta_{PLS} $ estimator requires a bias correction, just like ILS. The penalty term added to the SSR shrinks the factor $j$'s group of $N$ factor loadings, $z_{N,j}$, towards $0$. The adaptive group weights are based on the initial eigenvalue estimates, where small estimates, likely corresponding to redundant factors, increase the penalty. This process asymptotically separates the $m$ true factors from the $\hat m -m$ redundant ones. The PLS estimator requires $N$ and $T$ to be comparable, and it does not work with serially correlated errors, which means that such dynamics should be captured by the inclusion of lagged dependent variables.

\subsubsection{The Post Nuclear Norm Regularized Estimator}
The iteration in ILS begins with an initial $\beta$ and the next step is to obtain estimates for $F$ and $z_i$. Unfortunately, \cite{moonweidner2019} show that the minimization problem:
\begin{align}
	\underset{\{ z_{i}\}_{i=1}^N\in R^{m}, \{ f_{t}\}_{t=1}^T\in R^{m} }{min}SSR(\{ f_{t}\}_{t=1}^T, \{ z_{i}\}_{i=1}^N)&=\sum_{i=1}^N(y_i-X_i\beta-Fz_i)'(y_i-X_i\beta-Fz_i),\notag\\
 \text{subject to}\quad &m\leq \hat m.
\end{align} 
is not convex because the restriction $ m\leq \hat m $ is non-convex. This can lead the $ILS$, and also the $PLS$ due to its first step, iterative processes to converge to a local minimum or even to critical points which are not local minima. 

Brute-force computation can assist by conducting a grid search over the initial values of $\beta$, which serve as the starting point for the iterative estimation process. By evaluating the objective function across all grid points, one can plot its landscape and attempt to recover the global minimum. However, there is no guarantee of actually finding the global minimum, since this depends on whether the grid search has explored all relevant values of $\beta$-a task that is practically infeasible. A large $K$, or a large dataset, further burden the above approach.

\cite{moonweidner2019} suggest solving a convex minimization problem instead. This can happen by replacing the constraint $ m\leq \hat m $ with a restriction that is convex and which leads to an objective function which has a unique global minimum. Define $ \Gamma=zF' $, an $ N\times T $ matrix, where $ z=(z_1,z_2,...,z_N)' $ is a $ N\times m $ matrix. The restriction $ m\leq \hat m $ is equivalent to $ rank(\Gamma)\leq \hat m $. The convex relaxation of this constraint is $ \|\Gamma\|_1\leq C $ for some constant $ C $, where $ \|\cdot\|_1 $ is the nuclear or trace norm. The constrained minimization problem subject to $ \|\Gamma\|_1\leq C $ is equivalent to minimizing:
\begin{equation}
	Q_{\psi}(\beta)=\underset{\Gamma\in R^{N\times T}}{min}\left[ \frac{1}{2NT}\sum_{i=1}^N(y_i-X_i\beta-Fz_i)'(y_i-X_i\beta-Fz_i)+\frac{\psi}{\sqrt{NT}}\|\Gamma\|_1\right],
\end{equation} 
where $ \psi $ is a more convenient parameter instead of $ C $. \cite{moonweidner2019} propose two estimators which minimize the above objective function; 
\begin{equation}
	\hat\beta_{NNR,\psi}=\underset{\beta\in R^K}{armin}Q_{\psi}(\beta), \quad \text{and} \quad \hat\beta_{NNR,*}=lim_{\psi\to 0}\hat\beta_{NNR,\psi}.
\end{equation}
Both Nuclear-Norm Regularized (NNR) estimators are consistent but $\hat\beta_{NNR,*}$ has the advantage that it does not require specification of the nuisance parameter $\psi$, nor it does not require knowledge of $m$ or choosing a $\hat m$. It can be applied in the first step step of the iteration to get an estimate of $\beta$, and then the number of factors can be estimated in the next step. However, the price to pay is a reduced rate of convergence of marginally below $ \sqrt{min(N,T)} $ which is much slower than the $ \sqrt{NT} $ rate of the ILS estimator. For this reason, \cite{moonweidner2019} suggest the use of the Post Nuclear-Norm Regularized (PNNR) estimator which is $ \sqrt{NT} $ consistent. 

The PNNR estimator is described in the following steps:
\begin{enumerate}
\item For $s=0$, set $\hat\beta^{(s)}_{PNNR}=\hat\beta_{NNR,*}$ and the corresponding residuals $\hat u_{i}^{(0)}$.
\item Estimate the number of factors $\hat m $ from $\hat u_{i}^{(0)}$ using one of the methods for consistent estimation available to pure factor models, see e.g. \cite{Bai2002}, \cite{Onatski2010} and \cite{AhnHorenstein2013}.
\item Estimate the factor loadings $ \hat z_i^{(s+1)}$ and the factors $ \hat F^{(s+1)}$ by principal components, using $\hat m$ and $\hat u_{i}^{(s)}$.
\item Update the $\hat\beta^{(s)}_{PNNR}$ estimate using $ \hat F^{(s+1)}$ to:
\begin{equation}
\hat\beta^{(s+1)}_{PNNR}=\left(\sum_{i=1}^NX_i'M_{\hat F^{(s+1)}}X_i\right)^{-1}\left(\sum_{i=1}^N X_i'M_{\hat F^{(s+1)}} y_i\right),
\end{equation}
and obtain $\hat u_{i}^{(s+1)}$. Set $s=s+1$.
\item Iterate steps 3 and 4 a finite number of times. 
\end{enumerate}

The PNNR estimator is asymptotically equivalent to the combination of known number of factors and the ILS estimator. The same asymptotic bias corrections as in \cite{Bai2009} and \cite{moonweidner2017} can be applied to eliminate the bias of the PNNR estimator, for $s\geq 2$. The use of $\hat\beta_{NNR,\psi}$ in the PNNR estimator requires the assumption that $\hat m_{max}\geq m$ and an extra step for the estimation of $\psi$ from the data. However, these costs to be paid offer no benefit in terms of small sample performance as can be seen in the MC section of \cite{moonweidner2019}, except from a minimal bias-variance tradeoff. Overall, the PNNR estimator is attractive because it is based on the weakest possible assumptions and it does not suffer from many of the ILS estimators' shortcomings. 

\subsubsection{The Instrumental Variables Estimator}
The ILS and PNNR estimators suffer from asymptotic bias which, despite its correction in \eqref{eq:ilsbc}, can lead to poor small sample properties. If the regressors are strictly exogenous and $N\approx T$, \cite{CuiNorkuteSarafidisYamagata2022} suggest one-step and two-step instrumental variables (IV) estimators that have no asymptotic bias, and are also linear, avoiding the optimization difficulties in ILS or the computational burden of iteration in PNNR. These benefits come at the cost of two additional assumptions:

\begin{nassumption}[Linearity between regressors and factors]\label{ass:linear}
The interactive fixed effects are linearly related to the regressors:
\begin{equation}\label{eq:mundlak_IV}
X_i=F_xZ_{x,i}+\varepsilon_{x,i},
\end{equation}
where $Z_{x,i}$ is a $m_x\times K$ matrix of factor loadings and $\varepsilon_{x,i}$ is a $T\times K$ matrix of idiosyncratic terms.
\end{nassumption}

\begin{nassumption}[Independent factor loadings]\label{ass:indfactorload}
	The factor loadings $Z_{x,i}$ and $z_i$ are independent from $\varepsilon_{i}$, $\varepsilon_{x,i}$, $F_x$ and $F$. $Z_{x,i}$ and $z_i$ can still be correlated to each other.
\end{nassumption}

The estimation of \eqref{mod:ifemat}, \eqref{mod:ifemat2} and \eqref{eq:mundlak_IV} happens as follows. If $F_x$ is observed, the first step exploits the linearity assumption \eqref{eq:mundlak_IV} to defactor $X_i$: $\tilde{X}_i=M_{F_x}X_i=M_{F_x}\varepsilon_{x,i}$. The $\tilde{X}_i$ are no longer correlated with $u_i=Fz_i+\varepsilon_i$, since 
\begin{equation}
E[\varepsilon_{x,i}'M_{F_x}u_i]=E[\varepsilon_{x,i}'M_{F_x}Fz_i+\varepsilon_{x,i}'M_{F_x}\varepsilon_i]=0.
\end{equation}
The expected value of $0$ in the first term arises from the independent factor loadings assumption above, while the $0$ in the second term is obtained by regressor exogeneity, which implies that the idiosyncratic part of $X_i$, $\varepsilon_{x,i}$, is uncorrelated to the errors $\varepsilon_{i}$ - if that was the case additional external instruments would be necessary. 
Given that $\tilde{X}_i$ is correlated with $X_i$, through $\varepsilon_{x,i}$, and not to $u_i$, $\tilde{X}_i$ can be used as an instrument in the estimation of \eqref{mod:ifemat}. The first stage IV (FSIV) estimator is given by:
\begin{equation}\label{est:FSIV}
\hat\beta_{FSIV}=\left(\sum_{i=1}^N\tilde{X}_i'\tilde{X}_i\right)\sum_{i=1}^N\tilde{X}_i'y_i.
\end{equation}

The (infeasible) estimator $\hat\beta_{FSIV}$ is consistent for $\beta$, but asymptotically biased. Instead of correcting the bias, as with the ILS and PNNR estimators, \cite{CuiNorkuteSarafidisYamagata2022} add a second step in the IV estimation. Assuming now that $F$ is observed, the factor structures can be fully removed by pre-multiplying $X_i$ with $M_F M_{F_x}$ to get 
\begin{equation}
    \dbtilde X_i=M_FM_{F_x}=M_FM_{F_x}\varepsilon_{x,i}.
\end{equation}
The transformed regressor is  uncorrelated with $u_i$ since
\begin{equation}
E[\varepsilon_{x,i}'M_{F_x}M_{F}u_i]=E[\varepsilon_{x,i}'M_{F_x}M_{F}Fz_i+\varepsilon_{x,i}'M_{F_x}M_{F}\varepsilon_i]=0,
\end{equation}
but still correlated with $X_i$. Therefore, the second-step IV (TSIV) estimator of $\beta$ is:
\begin{equation}\label{est:TSIV}
\hat\beta_{TSIV}=\left(\sum_{i=1}^N\dbtilde{X}_i'\dbtilde{X}_i\right)\sum_{i=1}^N\dbtilde{X}_i' y_i.
\end{equation}

Both FSIV and TSIV estimators in \eqref{est:FSIV} and \eqref{est:TSIV} are infeasible in their present form as $F$ and $F_x$ are not observed. \cite{CuiNorkuteSarafidisYamagata2022} suggest estimating $m_x$ and $F_x$ from $X_i$ using principal components. Then, $\hat F_{x}$ should be used to get the feasible consistent estimator $\hat\beta_{FSIV}$ and the residuals $\hat{u}_i=y_i-X_i\hat\beta_{FSIV}$. Then, $m$ and $F$ can be estimated from $\hat u_i$ in the same way and used to obtain $\hat\beta_{TSIV}$. Obviously, the small sample properties of FSIV and TSIV depend on the quality of estimation of $m$, $m_x$, $F$ and $F_x$, as discussed in the analysis of the ILS estimator.

\cite{CuiNorkuteSarafidisYamagata2022} show that the feasible $\hat\beta_{TSIV}$ is asymptotically normal and has no bias when $T/N\to \rho$, with $0<\rho<\infty$
\begin{align}
    \sqrt{NT}(\hat\beta_{TSIV}-\beta)\xrightarrow{d}N(0,V).
\end{align}
In the special case where $F_x=F$ the FSIV estimator coincides with the ``PC'' estimator of \cite{westerlundurbain2015}, which requires bias correction. Additionally, the TSIV coincides with the FSIV. 

The assumption of linear dependence between regressors and individual effects was first introduced in \cite{mundlak1978} and has seen, in various forms, been used widely in panel data econometrics. As such we do not consider it as stringent. The factors in $X_i$, $F_x$ may or may not coincide with the factors in the interactive effects. If the two are completely independent then IFE do not pose an estimation problem as they are not correlated with the regressors. The more interesting and empirically plausible case therefore is when $F$ and $F_x$ are overlapping or are correlated. \cite{NorkuteEtal2021} impose no restrictions in the relationship between the two.

There is also no restriction in the relationship between $z_i$ and $Z_{x,i}$, which preserves the term ``fixed'' in ``interactive fixed effects'', as $X_i$ and $Fz_i$ can still be correlated through the permissible correlations between $F$ and $F_x$ and between $Z_{x,i}$ and $z_i$. If $f_t=f_{x,t}=1$ and $m=1$, \eqref{eq:mundlak_IV} coincides with the linearity assumption in \cite{mundlak1978}, which correlates regressors and fixed effects. 

Regarding the assumption that $F_x$ can be different to $F$, \cite{juodiskarawest2021} argue that such models arise naturally in the presence of lagged dependent variables in $ X_i $, because the additional factors in $X_i$ are linear combinations of the factors in $ F $, where the linear combination is constructed through backward substitution. To see this, consider the model $ y_{i,t}=\beta y_{i,t-1}+z_i f_t+\varepsilon_{i,t} $, where $ f_t $ is a scalar. Then, the representation of the regressor $ x_{i,t}=y_{i,t-1} $ is given by $  y_{i,t-1}=z_i(1-\beta L)^{-1}f_t+(1-\beta L)^{-1}\varepsilon_{i,t} $, where $ L $ is the backwards shift operator. The factor $ (1-\beta L)^{-1}f_t $ is an example of a factor which is a linear combination of a factor in $y_{i,t}$, but appears only in $ x_{i,t} $. 

The TSIV is applicable when the regressors include lagged dependent variables. In this case, \cite{NorkuteEtal2021} require the additional assumption:

\begin{nassumption}[Mean-zero and i.i.d factor loadings]\label{ass:indfactorloadTSIV}
The factor loadings are randomly distributed according to $Z_{x,i}\sim i.i.d.(0,\Sigma_{\Gamma_x})$ and $z_{i}\sim i.i.d.(0,\Sigma_{z})$. $Z_{x,i}$ and $z_i$ can still be correlated to each other.
\end{nassumption}

The assumption of mean-zero factor loadings simplifies the estimation problem but is likely not relevant in practice, as the zero mean imposes that asymptotically the factors are not relevant.

\subsubsection{Incidental Parameters-Free IV Estimator}
In the presence of lagged or endogenous variables, \cite{Juodis2022a} propose alternative, incidental parameters-free (IPF), IV and generalized method of moments (GMM) estimators that avoid estimating many nuisance parameters and thus are asymptotically unbiased. These estimators are applicable for any $T$, even when $T$ is fixed, making them an alternative to the Arellano and Bond GMM estimator \cite{arellanobond1991}, in the presence of IFE.\footnote{\cite{AHN2001219}, \cite{ahnlee2013panel} and \cite{Holtz-Eakin1988} examine IFE models in the same framework, however under more stringent conditions such as $m=1$.} The IPF estimators are based on an initial step of quasi-differencing that removes $z_i$, and then  a follow-up step that approximates $F$. 

For ease of exposition, assume that $m=1$ so that the factor $f_t$ is a scalar. The theory also assumes that $m$ is known. The quasi-differencing transformation of equations \eqref{eq:main} and \eqref{eq:fe2} removes $z_i$: 
\begin{equation}
    f_{t+1}(y_{i,t}-x_{i,t}'\beta)-f_{t}(y_{i,t+1}-x_{i,t+1}'\beta)=f_{t+1}\varepsilon_{i,t}-f_t\varepsilon_{i,t+1},\quad for \quad t=1,...,T-1,
\end{equation}
because $f_{t+1}z_if_t-f_tz_if_{t+1}=0$. This leads to the moment condition:
\begin{equation}\label{eq:ipfmom}
    m_{i,t}(\beta)=f_{t+1}(y_{i,t}-x_{i,t}'\beta)-f_t(y_{i,t+1}-x_{i,t+1}'\beta)\quad for \quad t=1,...,T-1.
\end{equation}
for which $E[m_{i,t}(\beta_0)|F]=0$, conditionally on the factors.

The moment \eqref{eq:ipfmom} cannot be used in estimating $\beta$ because $f_t$ is unobserved, and estimating $f_t$ for $t=1,...,T$ will lead to the incidental parameters problem as $T\to \infty$. Instead, \cite{Juodis2022a} suggest approximating $f_t$ via a proxy. Let $q_{i,t}$ be an instrumental variable, stochastic or non-stochastic, time-varying, which is correlated to $z_i$, but not to the errors:
\begin{nassumption}[Instrumental Variable]\label{ass:iv}
	\begin{equation}
    E[\varepsilon_{i,s}|q_{i,t},F]=0, \quad \text{and} \quad E[q_{i,t}z_i]\neq0.
\end{equation}
\end{nassumption}
The variable $q_{i,t}$ serves as the basis for creating weights that allow the approximation of $F$. Let $w_{i,t,s}=q_{i,t}d_{i,s}$, where $d_{i,s}=y_{i,s}-x_{i,s}'\beta=z_if_s+\varepsilon_{i,s}$. Then,
\begin{equation}\label{eq:weights}
    w_{t,s}=E[w_{i,ts}|F]=E[q_{i,t}z_i|F]f_s+E[q_{i,t}\varepsilon_{i,s}|F]=c_tf_s.
\end{equation}
Equation \eqref{eq:weights} says that the weights $w_{t,s}$ are linear combinations/proxies of the factors, and thus satisfy the moment condition in \eqref{eq:ipfmom} when inserted in the place of $f$:
\begin{equation}\label{eq:instfreemom2}
    m_{i,t}(\beta)=w_{t,t+1}(y_{i,t}-x_{i,t}'\beta)-w_{t,t}(y_{i,t+1}-x_{i,t+1}'\beta).
\end{equation}
The above moment conditions are nonlinear and require numerical optimization. \cite{Juodis2022a} additionally offer an alternative linear estimator under the Mundlak-type assumption that the factors enter linearly in the regressors. The number of factors needs to satisfy $m<T$, which is usually not an issue when $T$ is large. However, it must be estimated in a first step. \cite{Juodis2022a} demonstrate that $m$ can be consistently estimated, even for fixed $T$ by a BIC criterion.

\subsubsection{The Common Correlated Effects Estimator}
The common correlated effects (CCE) estimator proposed in \cite{Pesaran2006} was the first estimator developed for IFE. It's main strengths are simplicity in computation, the lack of need to estimate $m$, and good small sample properties. The CCE requires the assumptions of linearity between regressors and IFE and a rank condition to hold:

\begin{nassumption}[Linearity between regressors and factors]\label{ass:mundlakcce} 
The interactive fixed effects are linearly related to the regressors:
\begin{equation}
X_i=FZ_{x,i}+\varepsilon_{x,i},\label{mod:x}
\end{equation}
where $Z_{x,i}$ is a $m\times K$ matrix of factor loadings and $\varepsilon_{x,i}$ is a $T\times K$ matrix of idiosyncratic terms.
\end{nassumption}

\begin{nassumption}
	$m\leq K+1$.\footnote{ Technically the rank condition is on the loadings of the factors, which in practice boils down to the number of cross-section averages. Since we assume that all $K$ exogenous and additionally the dependent variable are used as cross-section averages, the rank condition implies that the number of factors must be equal to or less than the number of regressors plus one, i.e. $K+1$. This assumption is coupled with a regularity condition that the cross-section averages of the dependent and independent variables are informative for the factors \citep{Pesaran2006}.}
\end{nassumption} 

The key insight behind the CCEP estimator is that the cross-section-averages $\bar y=N^{-1}\sum_{i=1}^Ny_{i}$ and $\bar X=N^{-1}\sum_{i=1}^NX_{i}$ act as proxies for $F$, and thus can be added as additional regressors in \eqref{mod:ifemat}. \cite{Pesaran2006} showed: 
\begin{equation}
M_{\bar W}F=o_p(1)\label{eq:factdestr}
\end{equation}
where $\bar W=(\bar y,\bar X)$ and $M_{\bar W}=I_T-\bar W(\bar W'\bar W)^{-1}\bar W'$. The result in \eqref{eq:factdestr} says that the annihilator matrix $M_{\bar W}$ removes $F$ asymptotically, leading to a feasible version of the FE estimator in \eqref{est:ools}, named Pooled Common Correlated Effects Estimator (CCEP):
\begin{equation}\label{est:ccep}
	\hat \beta_{CCEP}=\left(\sum_{i=1}^NX_i'M_{\bar W}X_i\right)^{-1}\left(\sum_{i=1}^N X_i'M_{\bar W} y_i\right).
\end{equation}

The CCEP estimator does not require knowledge of the exact number of factors, subject to $m\leq K+1$. The errors $ \varepsilon_{i,t} $ can be serially or spatially correlated \citep{PesaranTosetti2011}, and the factors can be non-stationary when the regressors are strictly exogenous \citep{KPY2011}. Asymptotic inference is available for large $N$ and either fixed or large $T$, but there may be bias depending on the relative rate of expansion of $N$ and $T$. 

The most important assumption is the rank condition stating that the number of unobserved common factors must be equal to, or less, than the number of cross-section averages. The intuition is that the $ K+1 $ cross-section averages can be used to estimate up to $ K+1 $ unobserved factors. The rank condition has sparked considerable debate in the literature in the same way as $\hat m$ was important in ILS and PNNR estimation. Yet while there it was sufficient that $m\leq \hat m$, where $\hat m$ was determined by the researcher, the upper bound $K+1$ here is inflexible in that it depends on the model and the data. Recently, \cite{devossarafidis2024} propose a method for statistically testing if the rank condition holds. Nevertheless, the empirical factor literature, either in pure factor models or in panel data with IFE, usually finds only a small number of unobserved common factors. For example, \cite{robertson2018unit} examine Gibrat's law and find three factors in the errors. \cite{Juodis2022} examines the causes of increasing wage inequality and finds up to two factors stating ``Our procedure provides strong evidence that, irrespective of the setup considered, the number of the underlying factors is small in comparison with the total number of cross-section averages''. \cite{juodiskucinkas2023} find 1-3 factors in inflation expectations, and \cite{pesaransmith2023} find four factors in stock returns. \cite{mendenproano2017} find 3-4 factors in data on financial cycles based on 25 financial and 7 macro series for the US. A similar number of factors is found in the non-core bank funding rates of developed countries in \cite{CDK}. \cite{Gagliardini2019} find up to 5 factors in financial data and 5 factors in macro data. \cite{BoeseSchlosserEberhardt2024} find up to 4 factors in GDP data in 105 countries. \cite{devossarafidis2024} find up to 4 factors in the bank profitability of 450 banks.

A step that can be taken towards increasing the probability that the rank condition holds is to include any available observed common factors in $ \bar{W} $. Let $ d_t $ be a $ n\times 1 $ vector of $ n $ observed common factors and $ D=(d_1,d_2,...d_T)' $ a $ T\times n $ matrix stacking the factors over time, then the models \eqref{mod:ifemat} and \eqref{mod:x} become:
\begin{align}\label{mod:det}
	y_i&=D a_i+X_i\beta_i+u_i,\\
	X_i&=D A_i+FZ_{x,i}+\varepsilon_{x,i},
\end{align}
where $ a_i $ is a $ n\times 1 $ vector and $ A_i $ is $ T\times n $ matrix of observed common factor loadings. Estimation proceeds as previously except that $ \bar{W}=(D,\bar{y},\bar{X}) $. The above formulation assumes that there are $ n $ observed common factors in $ D $ and $ m $ unobserved ones in $ F $. The rank condition remains the same in that it must be $ m\leq K+1 $. Therefore any observed common factors may be taken out of $ F $ and into $ D $ to reduce $ m $. Examples of such factors are the individual effects, seasonal dummies, average stock returns, GDP growth, the VIX volatility index, and oil prices among others. 

A final step in this direction would be to employ $ \bar{W}=(D,\bar{y},\bar{X},\bar{X}^*) $, where $ \bar{X}^* $ are cross-section averages of covariates which are affected by the same set of common factors but are not included in the regression  \citep{ChudikPesaran2015}. If the number of these covariates is $ K^* $, then the rank condition becomes $ m\leq K+K^*+1  $. In practice, this is reasonable in many settings. In the forecasting literature, \cite{stockwatson2002} and \cite{forni2005}, assume that there exist ``global'' underlying factors such as technology or general economic conditions that affect many variables beyond those directly involved in the relationship of interest. 

Finally, \cite{ChudikPesaranTosetti2011} show that CCEP remains consistent and asymptotically normal if in addition to the $m$ strong factors there are additional $m_w$ weak factors which however, are not correlated with the regressors, and $ m_w\to \infty $. Weak factors can capture spatial dependence and thus CCE can be used, as long as that spatial dependence is not correlated to the regressors.

While the rank condition is necessary for consistency, bias arises when there are more cross-section averages than factors, a problem which is exacerbated in dynamic panels. If regressors in $ x_{i,t} $ are strictly exogenous to the error term $ \varepsilon_{i,t} $. \cite{karabiyik2017} show that:
\begin{equation}
	\sqrt{NT}(\hat\beta_{CCEP}-\beta)\xrightarrow{d}N(0,V)+\sqrt{\rho}(B_1-B_2),
\end{equation} 
where $ V$ is the asymptotic variance of $ \hat\beta_{CCEP} $, $ T/N\to \rho $, and $ B_1 $, $ B_2 $ are bias terms. If $ \rho =0 $, then the CCEP estimator is asymptotically unbiased. Alternatively, if $ \rho>0 $ the bias is a function of $ B_1-B_2 $, where $ B_1 $ is the bias arising from the need to estimate the factors \citep{westerlundurbain2015}, and $ B_2 $ is the bias of having more cross-section averages than factors, a term which appears only when $m<K+1$, but not $m=K+1$ \citep{karabiyik2017}. \cite{DeVos2024} offer a bootstrap methodology that removes the bias from CCEP.

If the model includes lagged dependent variables the factor space can be approximated by augmenting the annihilator matrix with a finite number of lagged cross-section averages of order $ p_T=[T^{1/3}] $ \citep{ChudikPesaranJoE2015,devoseveraert2021}. The lagged dependent variables also introduce a Nickell bias to the estimator:
\begin{align}
	\sqrt{NT}(\hat\delta_{CCEP}-\delta)\xrightarrow{d}N(0,V)+O_p\left(\frac{N}{T}\right)+O_p\left(\frac{T}{N}\right),
\end{align} 
where the Nickell bias is the $O_p(N/T)$ term and $B_1$ and $B_2$ are included in the $O_p(T/N)$ term. Since it cannot be that both $ T/N\to 0 $ and $ N/T\to 0 $, bias correction is crucial to make CCEP usable in this case. \cite{devoseveraert2021} suggest an effective bias correction for the Nickell bias $ O_p\left(N/T\right)$ making CCEP applicable to dynamic models. The bias-corrected CCEP requires lags of the cross-section averages as above, but only as many as the autoregressive order of $ x_{i,t} $, assuming that $ x_{i,t} $ follows an autoregressive process. Yet in practice this order is rarely known so \cite{devoseveraert2021} conclude that lags of order $ p_T=[T^{1/3}] $ should be used as a precaution against misspecification. Their Monte Carlo simulations show that additional lags have a diminishing impact on estimator performance as $ T $ grows.

In dynamic panels estimating more factors than the true number can severely impact the estimator's rate of convergence. \cite{juodiskarawest2021} consider the model:
\begin{align}
	y_{i}&=X_{i}\beta+F_y z_i+\varepsilon_i\\
	X_i&=F_yZ_{y,i}+F_{-y}Z_{-y,i}+\varepsilon_{x,i},
\end{align}  
where $ F_y $ is the set of $ m_y $ factors, or ``y-factors'', and $ F_{-y} $ is the set of $ m-m_y $ additional factors which appear in $ X_i $ only, as in \cite{CuiNorkuteSarafidisYamagata2022}. This last set of factors appears naturally in models with lagged dependent variables. It turns out that the $ K+1 $ cross-section averages in $ \bar {W} $ can estimate the factor space of the $ m_{mean} $ factors $ F_{mean}=(F_y,F_{-y}R) $. $ F_{mean} $ includes $ F_y $, plus a linear combination of $ m_{mean}-m_y $ factors in $ F_{-y} $. The remaining $ m-m_{mean} $ factors $ F_{-mean} $ are simply the residuals from projecting out $ F_{mean} $ from $ F_{-y} $ and these cannot be estimated by cross-section averages. Any mean factors must be strong factors, while non-mean factors can be either strong or weak. The main finding is that, if $ K+1>m_{mean}$ and $m>m_{mean}$, with no restrictions on the relationship between $K+1$ and $m$, then the CCE estimator is $ \sqrt{N} $ consistent but not asymptotically normal. This is an empirically plausible scenario where the number of regressors are more than the important mean factors, resulting in reduced convergence rate as can be seen in the expansion:
\begin{equation}
	\sqrt{NT}(\hat\beta_{CCEP}-\beta)=\Sigma_X^{-1}\left(b_0+\sqrt{\frac{N}{T}}b_1+\sqrt{\frac{T}{N}}b_2+\sqrt{T}\xi\right)+o_p(1),
\end{equation}  
where $ b_0 $ is a mean-zero asymptotically normal random variable, $ b_1 $ is the Nickell bias arising from the inclusion of lagged dependent variables in the model, $ b_2 $ is the incidental parameter bias arising from factor estimation, and $ \xi $ is an $ O_p(1) $ non-linear function of the factor loadings of the $ F_{-y} $ factors. This last term is the reason for the loss of the faster rate of convergence and for the non-standard asymptotic distribution, due to which inference is no longer possible. 

Overall, the benefit of not having to estimate the number of factors comes at the cost of a bias when $K+1>m$. The literature proceeds with methods for selecting a number of cross-section averages $K^\star=m$ to remove the bias. \cite{Juodis2022} proposed a regularization method to deal with the case if $ K+1>m_{mean} $ by estimating the correct number of factors in a first step. The regularised CCE estimator involves the use of the Eigenvalue Ratio (ER) approach of \cite{AhnHorenstein2013} to estimate, first the number of factors $m_y$ in $ F_y $, based on the eigenvalues of the cross-section averages in $ \bar W $, and secondly $ \hat F_y $ based on the eigenvectors related to the $\hat m_y $ greatest eigenvalues. Then, CCEP is given by \eqref{est:ools} with $\hat F_y$ in place of $F$. This approach restores the asymptotic normality of the CCE estimator, however a bootstrap scheme is needed for inference. \cite{LucaMargaritella2023} propose the application of the Akaike and Schwarz Bayesian information criteria on the cross-section averages to estimate the number of factors and identify the optimal set of cross-section averages. This method can also be employed when the available variables for the cross-section averages come from an expanded list which includes observed factors, or covariates not included in the regression. However, the impact of this first-step on the subsequent CCEP estimation is not studied. \cite{DeVos2024} propose an alternative version based solely on $X_i$, which is applicable even when there are unique factors in $X_i$. They argue that inference can proceed using the proposed bootstrap. \cite{DitzenStauskas2025} point out that the criteria in both cases should only be used for static panels and stationary factors.

\subsubsection{Time-Averaged CCE}
\cite{westerlundkadkara2025} propose a version for CCE applicable to fixed $N$ panels or SUR models with IFE in the errors. The idea is to use the time averages $\bar y_i^T=T^{-1}\sum_{t=1}^Ty_{i,t}$ and $\bar x_i^T=T^{-1}\sum_{t=1}^Tx_{i,t}$ instead of cross-section averages to remove the loadings $z_i$, and hence the new estimator is termed Time-Averaged CCE (TACCE). The assumptions of TACCE are similar to CCEP, including a rank condition.

Assuming that $N$ is fixed is advantageous in that it makes the results of this paper applicable in panels which small $N$ dimension such as portfolios, commodities, industries, regions and countries. The new methods are uniquely able to shed light in small groups of individuals or aggregates that previously could not be properly examined. In growth econometrics there are many such examples, such as the original $20$ OECD countries, the $20$ Eurozone countries, the $10$ ASEAN countries, the $G7$, the $20$ MENA countries, the $10$ Canadian provinces and the
$4$ UK nations among many other groupings. Fixed $N$ analysis is especially useful in robustness analyses, where typically a sample is split into sub-samples and their subsections typically estimated in robustness analysis. Finally, the fixed $N$ analysis may provide more complete results because it allows preservation of the historical information. Many panel studies consider shorter periods of observation in order to increase the number of units in the sample. By doing so, the estimation results reflect the arbitrariness of the selected time period rather than long-run relationships. 

\subsection{Estimators for Panel Data with Grouped Fixed Effects} \label{sect:GFE}
As can be seen from the previous section, the IFE model estimation is accompanied with the incidental parameters problem which introduces bias to the slope coefficients. Grouped fixed effects were proposed in \cite{Bonhomme2015a} as a way to reduce the number of parameters that needed to be estimated. GFE offers parsimony under the assumption that units belong to a small number of groups. 

\subsubsection{The GF estimator}
Let the number of groups $G$ be known and denote group membership by $ g_i=g $, which means that unit $i$ belongs to group $g$. The the Grouped Fixed Effects (GF) estimator minimizes:
\begin{equation}
\left(\hat \beta,\hat c, \hat \gamma\right)=\underset{(\beta,c,\gamma)\in R^K\times R^{GT}\times \Gamma_G} {argmin} \sum_{i=1}^N\sum_{t=1}^T\left(y_{i,t}-x_{i,t}'\beta-c_{g_i,t}\right)^2,
\end{equation}
where $c$ denotes all the group-specific time effects and $\gamma=\{g_1,...,g_N\}$ is a particular grouping of the $N$ units into $G$ groups, belonging to the set of all possible groupings $\Gamma_G$. The individual effect $ c_{i,t} $ will be denoted in this section by $ c_{g_i,t} $ as it varies across groups and not units.

The above minimization problem can be solved iteratively as with the ILS estimator. There are two steps here as well, where the second step involves group allocation instead of estimating the factors. Given some initial values for $\beta$ and $c$, each individual's group membership can be determined as:
\begin{equation}
\hat{g}_i(\beta,c)=\underset{g\in\{1,2,...,G\}}{argmin}\sum_{t=1}^T \left(y_{i,t}-x_{i,t}'\beta-c_{i,t}\right)^2,
\end{equation}
by the k-means algorithm. Notice that group membership is determined solely through the time dimension. Once group membership has been estimated, estimates of the slope coefficients and the grouped fixed effects can be calculated as:
\begin{equation}\label{eq:gfobject}
\left(\hat \beta,\hat c\right)=\underset{(\beta,c)\in R^K\times R^{GT}} {argmin} \sum_{i=1}^N\sum_{t=1}^T\left(y_{i,t}-x_{i,t}'\beta-c_{\hat{g}_i(\beta,c),t}\right)^2.
\end{equation}
The difficulties in applying the GF estimator are similar to those for the ILS estimator. The GF estimator asymptotic theory is derived under the assumption that the number of groups is known. If $ G $ is unknown, \cite{Bonhomme2015a} suggest estimating $ G $ with the BIC criterion.\footnote{\cite{suju2018} propose estimating the number of factors and the number of groups using BIC-type information criteria.} The GF estimators of both $\hat \beta$ and $\hat c_{g_i,t}$ are asymptotically normal with $\sqrt{NT}$ and $\sqrt{N}$ rates of convergence respectively.

The argument for applying the GF estimator is parsimony and improved small sample performance. Group fixed effects represent a restriction of interactive fixed effects, and this restriction leads to efficiency gains when it holds, and inconsistency for the GF when it does not hold. The following assumptions are necessary:
\begin{nassumption}
The groups are well separated: for all $(g,\tilde g)\in\{1,...,G\}^2$ such that $g\neq \tilde g$:
        \begin{equation}
            \underset{T\to\infty}{plim}\frac{1}{T}\sum_{t=1}^{T}(c_{g,t}-c_{\tilde g,t})^2=b_{g,\tilde g}>0.
        \end{equation}
\end{nassumption}
\begin{nassumption}
Error and regressor distribution tails decay at a faster-than-polynomial rate.
\end{nassumption}

The first assumption can be violated if two groups are too close to each other.\footnote{This case has not been studied in the literature yet. It may be interesting to see what happens near-violation with $b_{g,\tilde g}=b/T$.} The second assumption requires distributions with fast tail decay, so that large deviations from the mean are unlikely. Permitted distributions include sub-gaussian and sub-exponential distributions, as well as distributions without tails such as the uniform and Bernoulli. This assumption can be violated by outliers. Under the above assumptions, the group misclassification probability decays to zero exponentially as $ T $ increases \citep{Bonhomme2015a}, making the estimator useful even in scenarios where the time dimension is not large:
\begin{equation}
 	Pr\left(\underset{i\in\{1,...,N\}}{sup}\left\|\hat g_i-g_i\right\|>0\right)=o(1)+o\left(NT^{-\delta}\right),
\end{equation}
where $ \delta>0 $ and $ N $, $ T $ diverge to infinity. For $ \delta>1 $ the time dimension can be much smaller than $ N $. A key impact of the assumption of group membership is that the number of groups is fixed and does not increase as $ N $ increases. This assumption removes the incidental parameter problem, and with it, both the $ O(1/N) $ bias and the $O(1/T)$ Nickell bias which appears if $ x_{i,t} $ includes lagged dependent variables.

\cite{Bonhomme2015a} suggest two algorithms for calculating the GF estimator, however both of them are computationally intensive. The first is the k-means algorithm of \cite{forgy1965}. To deal with the existence of local minima, it is suggested that many different starting values for $\beta$ and $c$ should be used in order to make sure that the solution is the lowest value of the objective function \eqref{eq:gfobject}. The accompanying software suggests at least $1,000,000$ different starting values. \cite{manresachetverikov2022} use only 30, standard normal $R^{K+G\times T}$, starting values in their MC experiment as they state ``because even with this number calculating the grouped fixed-effect estimator takes non-trivial time''. \cite{mugnier2025} reports that the clustering algorithm breaks down when too many starting values are drawn close to 0, as when drawn from a normal distribution. This situation becomes worse as $G$ and $T$ increase. The second algorithm is based on Variable Neighborhood Search and is more efficient, but requires the specification of two more tuning parameters which can affect its performance.  

The above discussion hints at the difficulties of using the k-means algorithm. First, the k-means algorithm solves an NP-hard problem which means that any fast algorithm, such as the Variable Neighborhood Search, must occasionally fail to reach the true values. The alternative of searching using many different starting values for $\beta$ and $c$ has the weakness that there is no way to know if the global minimum has been reached, unless all possible partitions of units into groups has been tried. This number is huge: $G^N$. Third, it requires that the number of groups or at least an upper bound to this number is known. All these difficulties can be avoided if GFE is treated as IFE and an IFE estimator is used, however, the problem in this case is that there is no parametric rate inference \citep{moonweidner2019,manresachetverikov2022}. The reduction to the convergence rate is analogous to what happened to the CCEP estimator when the cross-section averages were more than the number of factors.

\subsubsection{The Post-Spectral Estimator}
\cite{manresachetverikov2022} provide an alternative estimator that is computationally straightforward and does not employ the k-means algorithm, under the Mundlak-type assumption which imposes a linear relationship between the regressors and the grouped fixed effects:
\begin{nassumption}
    The regressors are generated as:
    \begin{equation}
        x_{i,t}=\sum_{d=1}^Dz_{x,i,d}'c_{g_i,t}^d+\varepsilon_{x,i,t},
    \end{equation}
    for $d=1,...,D$.
\end{nassumption}
The grouped fixed effects from the $y$-equation of the model are those for $d=1$, while for $d=2,...,D$ there are other grouped fixed effects which are specific to $x_{i,t}$, just like the sets of factors $F_{y}$ and $F_{-y}$ in CCEP. Further necessary assumptions include that the groups are well separated and that the errors are sub-gaussianly distributed - an assumption slightly stronger than the one used by \cite{Bonhomme2015a}.

The estimation method proceeds in three steps:
\begin{enumerate}
    \item An initial estimator for $\beta$. Define the symmetric $N\times N$ matrix of unit error distances $A^\beta$ with elements:
    \begin{equation}
        A^\beta_{i,j}=\frac{1}{NT}\sum_{t=1}^T\left[\left(y_{i,t}-x_{i,t}'\beta\right)-\left(y_{j,t}-x_{j,t}'\beta\right)\right]^2
    \end{equation}
    for all $i,j=1,...,N$. Then, for the $2GD+2$ largest eigenvalues $\lambda^\beta_{1},...,\lambda^\beta_{2GD+2}$ it holds that:
    \begin{equation}
    \lambda^\beta_{1},...,\lambda^\beta_{2GD+2}=\beta'\Sigma\beta+S'\beta+L+o_{p}(1),
    \end{equation}
    where $\Sigma$ is symmetric and positive definite. The initial spectral estimator is the minimizer of the above convex function:
    \begin{equation}
        \tilde\beta_{SP}=arg\underset{\beta\in R^K}{min}\beta'\hat\Sigma\beta+\hat S\beta+\hat L,
    \end{equation}
    where $\Sigma$, $S$ and $L$ are estimated as functions from the eigenvalues of $A^\beta$. The spectral estimator $\tilde\beta_{SP}$ solves a convex optimization problem and is an alternative to the NNR estimator of \cite{moonweidner2019} with a faster rate of convergence $min\{T,N\}$. However, \cite{manresachetverikov2022} show that the NNR estimator can also have an $min\{T,N\}$ rate of convergence under a Mundlak-type linearity condition.
    \item Classification of units into groups. The classification algorithm is based on asymptotic principal components as in \cite{Bai2002}. It does not solve any non-convex optimization problems as the k-means algorithm does, and therefore does not suffer from any of its drawbacks.
    \item Post-Spectral (PS) Estimator. Given that the group membership has been estimated, the PS estimator is simply pooled OLS on that model:
    \begin{equation}
        \left(\beta_{PS},\hat c\right)=\underset{(\beta,c)\in R^K\times R^{GT}}{argmin}\sum_{i=1}^N\sum_{t=1}^T(y_{i,t}-x_{i,t}'\beta-c_{\hat g_i,t})^2
    \end{equation}
\end{enumerate}
An additional advantage of the PS estimator over GF is that it requires a much smaller $T$ than GF, although this theroetical advantage does not materialize in the Monte Carlo simulations of \cite{mugnier2025}, where it seems that the GF estimator dominates for all $N$ and $T$ in terms of bias and RMSE. In that experiment, for $N=180$, $T=40$ and $G=4$, the GF estimator however, takes $1.55$ seconds to compute while the PS only $0.87$. \cite{manresachetverikov2022} demonstrate how the PS estimator can be applied to dynamic and high-dimensional panels, and to panels with interactive fixed effects as a first step to ILS. 

\subsubsection{The Three-step Triad Pairwise-Differencing Estimator}
\cite{mugnier2025} proposes another alternative estimation methodology which attempts to remove the computational burden of the GF estimator. At the same time however, it addresses a weakness in the PS and GF estimators that either $G$, or an upper bound to $G$, is known. While the former is rarely possible in practice, the latter leads to the estimation of multiple misspecified models as part of the calculation of a BIC-type criterion. Instead, this approach estimates the number of groups at the same time it estimates group membership. Finally, the new method does not require a Mundlak-type assumption on the regressors, as the PS estimator. The resulting procedure, summarized below, has polynomial time complexity:
\begin{enumerate}
    \item An initial estimator for $\beta$. Estimate $\beta$ with an initial estimator $\hat\beta^1$ which solves a convex constrained optimization problem. This could be the NNR estimator from \cite{moonweidner2019}, and preferably $\hat\beta_{NNR,*}$ which has no tuning parameter.
    \item Classification of units into groups. Apply agglomerative clustering based on distance thresholding. This is a bottom-up approach which starts with $N$ singleton groups and creates clusters by looking at pairwise distances across unit residual correlations $\hat v_{i,t}$, where $\hat v_{i,t}=y_{i,t}-x_{i,t}'\hat\beta^1$ and the distance is:
    \begin{equation}
        \hat d^2_{\infty,1}=\underset{k\in\{1,...,N\}/ \{i,j\}}{max} \left|\frac{1}{T}\sum_{t=1}^T\left(\hat u_{i,t}-\hat v_{j,t}\right)\hat v_{k,t}
        \right|.
    \end{equation}
    Asymptotically, the distance between two units in the same group should be zero, as can be seen by:
    \begin{equation}
        \frac{1}{T}\sum_{t=1}^T\left(\hat u_{i,t}-\hat v_{j,t}\right)\hat v_{k,t}\approx \frac{1}{T}\sum_{t=1}^T\left(\alpha_{g_i,t}-\alpha_{g_j,t}\right)\alpha_{g_k,t}.
    \end{equation}
    The algorithm proceeds to add to pairs of units that belong in the same group other units in the vicinity. The outcome of this process is the identification of the group structure in the model.
    \item The Three-step Triad Pairwise-Differencing (TPWD) Estimator. Compute the pooled OLS estimator with interactions of time and estimated group dummies. Use the OLS estimator standard errors for inference.
\end{enumerate}
Based on Monte Carlo simulations, \cite{mugnier2025} argue that an iteration of this scheme, where the final TPWD estimator is employed as the initial estimator in the first step and repeating until convergence, leads to notable improvements in terms of bias and RMSE. The TPWD estimator can be applied to models with exogenous or predetermined regressors and under weak assumptions on the errors. Monte Carlo simulations suggest that it outperforms in terms of bias, RMSE and coverage other existing estimators such as PS, but not GF, which however is more costly to compute. In terms of computational burden, the challenge is not totally removed. Simulations in \cite{mugnier2025} demonstrate that doubling $N$ leads to a 10-fold increase in computational time. This is likely due to the memory required in storing the differences of the residuals at the clustering step. Furthermore, the $\hat{G}$ estimation requires significantly larger $T$ than the BIC-type criteria.  

\subsection{Estimators for Panel Data with Nonseperable Unobserved Heterogeneity}\label{sect:NSTW}
The NSTW literature was born out of advances in the estimation of GFE models. It became apparent that the clustering methods employed to capture group structures can be used to capture group structures also in more complex forms of unobserved heterogeneity, in such way that the remaining errors, arising from the nonexistence of actual groups, are asymptotically trivial.

\subsubsection{The Two-Step GF Estimator}
\cite{bonhommelamadonmanresa2022} introduce the idea of employing the discrete heterogeneity methods applied in the grouped fixed effects literature to approximate general forms of continuous heterogeneity. When $h(\cdot,\cdot)$ in $c_{i,t}=h(z_i,f_t)$ is additive (TWFE) or multiplicative (IFE), the estimators of section \ref{sect:IFE} can be used. However, if $h(\cdot,\cdot)$ is nonlinear, this is no longer the case. Discretization involves clustering individuals and time periods to a smaller number of groups, based on observed variable averages which are assumed to be informative about $z_i$ (injectivity assumption). These averages are only an approximation to $h(\cdot,\cdot)$, but the number of groups can be selected in such way so that the approximation errors are asymptotically trivial. To avoid introducing burdensome notation we present the injectivity assumption informally:
\begin{nassumption}
    Two individuals with the same model population moments have the same $z_i$ . 
\end{nassumption}
Additionally, they employ a relaxed Mundlak-type assumption: 
\begin{nassumption}\label{ass:nstw-lin}
    $x_{i,t,k}=g_k(z_i,f_t)+\varepsilon_{i,t,k},$
\end{nassumption}
\noindent where $g(\cdot,\cdot)$ is unknown and does not need to be specified. The latter assumption justifies using $x_{i,t}$ in the classification step below. The exposition in \cite{bonhommelamadonmanresa2022} focuses on general moment conditions, but in the following we will simplify it, adapting it to the normal linear regression \eqref{eq:ie}. Define $\xi_i=T^{-1}\sum_{t=1}^T(x_{i,t}',y_{i,t})'$ for $i=1,...,N$ and $\psi_t=N^{-1}\sum_{i=1}^N(x_{i,t}',y_{i,t})'$ for $t=1,...,T$.
\begin{enumerate}
    \item Classification. The first step is to employ k-means to partition individuals into $G$ groups across the cross-section dimension: 
    \begin{equation}
        \left(\hat \xi(1),....,\hat \xi(G), \hat g_1,..., \hat g_N \right)=\underset{ \xi(1),...,\xi(G), g_1,...,g_N }{argmin} \sum_{i=1}^N \left\|\xi_i-\xi_i (g_i) \right\|^2,
    \end{equation}
    and into $C$ groups across the time dimension:
    \begin{equation}
        \left(\hat \psi(1),....,\hat \psi(C), \hat l_1,..., \hat l_T \right)=\underset{ \psi(1),...,\psi(C), l_1,...,l_T }{argmin} \sum_{t=1}^T \left\|\psi_t-\psi_t (l_t) \right\|^2.
    \end{equation}
    \item Estimation. Given group membership, the second step is to maximize the model's log-likelihood function with respect to the slope coefficients and the group fixed effects. Assuming normal and homoskedastic errors, the two-step GF (TSGF) estimator is:
    \begin{equation}\label{eq:blm_mod}
        \hat \beta_{TWGF}=\underset{\beta\in R^K}{argmin}\sum_{i=1}^N\sum_{t=1}^T(y_{i,t}-x_{i,t}'\beta-c_{g,l})^2,
    \end{equation}
    where $c_{g,l}$ is the fixed effect specific to each combination of groups $(g,l)\in \left\{1,...,G\right\}\times \left\{1,...,C\right\}$.
\end{enumerate}
Under the injectivity assumption, and the assumptions of a static model and of independent observations across both dimensions, a Taylor rule shows that:
\begin{equation}\label{eq:tsgf_bias}
    \hat \beta=\beta+H^{-1}\frac{1}{N}\sum_{i=1}^Ns_i+O_p\left(\frac{1}{T}+\frac{1}{N}+\frac{GC}{NT}\right)+O_p\left(G^{-2/K}+C^{-2/K}\right)+o_p\left(\frac{1}{\sqrt{NT}}\right).
\end{equation}
The $O_p(T^{-1})$ and $O_p(N^{-1})$ biases are incidental parameter appearing also in the fixed effects estimators. Here they appear due to classification noise from the first step of the k-means algorithm. The $O_p((NT)^{-1}GC))$ bias arises from the estimation of $GC$ group-specific parameters using $NT$ observations. 

The remaining $O_p\left(G^{-2/K}+C^{-2/K}\right)$ bias comes from the approximation error from applying the k-means algorithm when heterogeneity does not really have group structure. These terms can go away by appropriately selecting the number of groups: 
\begin{align}
    \hat G&=\underset{G\geq 1}{min}\left\{G: \hat Q_g(G)\leq \gamma \hat V_g\right\},\\
    \hat C&=\underset{C\geq 1}{min}\left\{C: \hat Q_c(C)\leq \gamma \hat V_c\right\},
\end{align}
where $\gamma\in(0,1]$ is a user-specified parameter and $\hat Q_g(G)=N^{-1}\sum_{i=1}^N\left\|\xi_i-\hat \xi( g_i) \right\|^2$ and $\hat Q_c(C)=T^{-1}\sum_{t=1}^T\left\|\psi_t-\hat \psi(l_t) \right\|^2$ are the k-means objective functions. $V_g$ and $V_c$ capture noise and can be estimated in various ways. A simple formula for independent in time errors is: $\hat V_g=(NT^2)^{-1}\sum_{i=1}^N\sum_{t=1}^T\left\|\xi_{i,t}-\hat \xi_i \right\|^2$ and $\hat V_c=(TN^2)^{-1}\sum_{t=1}^T\sum_{t=1}^T\left\|\psi_{i,t}-\hat \psi_t \right\|^2$.

The TSGF estimator is consistent, but no asymptotic distribution theory is available for it. The rate of convergence is not fast either, as can be seen from \eqref{eq:tsgf_bias}. \cite{bonhommelamadonmanresa2022} argue that if $m=1$, again for the static model with independent observations across units and time, the rate of convergence is $min\{N,T\}$. Asymptotic inference becomes possible for a variant of this estimator, in the extension by \cite{Beyhum2024} presented below. 

\subsubsection{The ILS and TSGF-H Estimators}
\cite{FREEMAN2023105498} make the point that the NSTW model of unobserved heterogeneity of \cite{bonhommelamadonmanresa2022} can be seen as an approximate IFE model with some strong factors, and an infinite number of weak factors, all of which allowed to be correlated to the regressors. As such, the ILS estimator is still applicable under NSTW, yet the number of factors need to grow asymptotically to approximate $h(z_i,f_t)$. The caveat is that the rate of convergence is slow and equal to $min\{\sqrt{N},\sqrt{T}\}$, however, simulation evidence in Freeman and Weidner (2023) shows that the estimator performs well and that the slow convergence rate could be an artifact of the proofs.

\cite{FREEMAN2023105498} develop one more estimator, called TSGF-H here, where ``H'' stands for hierarchical clustering. TSGF-H has improved small sample performance compared to ILS by discretizing unobserved heterogeneity in a first step. Unlike \cite{bonhommelamadonmanresa2022}, injectivity is assumed for both $z_i$ and $f_t$ and there is no need for the Mundlak-type assumption. In the first step, hierarchical clustering is employed instead of k-means, and the second step is also different, employing TWFE in:
\begin{equation}\label{eq:fw_mod}
y_{i,t}=x_{i,t}'\beta+\delta_{i,l_t}+\nu_{t,g_i}+\varepsilon_{i,t}. 
\end{equation}
Compare the difference between models \eqref{eq:fw_mod} and \eqref{eq:blm_mod}; the individual effects for each unit $i$ are preserved, yet they are allowed to vary in time across the different time-groups. A similar condition applies for the time effects. The latter model can be justified by the fact that the unobserved heterogeneity function $h(z_i,f_t)$, which is approximated by $\delta_{i,l_t}$ and $\nu_{t,g_i}$ in \eqref{eq:fw_mod}, can be expanded as:
\begin{equation}
h(z_i,f_t)=\delta_{i,l_t}+\nu_{t,g_i}+O\left(\|z_i-\bar z_{g_i}\|^2+\|f_t-\bar f_{c_t}\|^2
    \right),
\end{equation}
where the approximation errors around the group means $\bar z_{g_i}$ and $\bar f_{c_t}$ are quadratic, compared to the linear errors arising from model \eqref{eq:blm_mod}, where $h(z_i,f_t)$ is approximated by $c_{g,l}$: 
\begin{equation}
h(z_i,f_t)=h(\bar z_{g_i},\bar f_{c_t})+O\left(\|z_i-\bar z_{g_i}\|+\|f_t-\bar f_{c_t}\|
    \right).
\end{equation}
The hierarchical clustering algorithm allows for groups of size 2-3 so that the groups are less than $N$ in order to avoid the incidental parameter bias but enough that the unit parameters within each group are close enough. The algorithm does not require a specification of $G$ and $C$ as in k-means. Instead, it depends on the choice of a large $m$ in the first step and then on the choice of a truncated $m^\star$, which is the number of estimated factor loadings and factors on which the clustering is based on. \cite{FREEMAN2023105498} derive consistency and the rate of convergence, but not the TSGF-H's asymptotic distribution.   

\subsubsection{The TSGF-KT Estimator}
\cite{Beyhum2024} combine the first-stage k-means clustering approach of \cite{bonhommelamadonmanresa2022} with the TWFE estimator in the second step of \cite{FREEMAN2023105498}. That is they employ the additively separable two-way group fixed effects in equation \eqref{eq:fw_mod} instead of the nonseparable two-way group fixed effects, specific to the intersection of unit and time clusters in \eqref{eq:blm_mod}. The above step combination finally allows the derivation of the asymptotic distribution of the TSGF estimator in the last step of the procedure. We name this estimator TSGF-KT where ``K'' stands from the first-step k-means clustering and ``T'' stands for the two-way group fixed effects in the second step.

The asymptotic distribution is derived on the observation that the second-step estimator in \eqref{eq:fw_mod} is based on a Neyman-orthogonal moment. This moment is different to that employed in the second step of \cite{FREEMAN2023105498}, because of the different first step. Neyman-orthogonal moments reduce bias; they ensure that errors made when estimating the nuisance parameters in the first step have only limited influence on the final estimator and hence, valid inference remains possible even when nuisance parameters are estimated. 

The TSGF-KT estimator has a $\sqrt{NT}$ rate of convergence and is asymptotically normal, with the usual standard errors still valid. Monte Carlo simulations in \cite{Beyhum2024} show that the estimator performs well for $T$ as small as $10$ and $N=50$. 

\section{Behind the Estimates: Detecting Misspecification in Panel Models}
In this section we discuss methods and tests which guide the application of the estimators in the previous section. The existence of IFE, GFE and NSTW implies that the cross-section dependence across units is strong, and therefore, the existence of these forms of unobserved heterogeneity is often tested by tests of cross-section dependence \citep{SarafidisWansbeek2012}. 

Another way to detect more complex structures of heterogeneity is by estimating the number of strong factors in the variables and in the residuals. Therefore, the methods of estimating $m$ have a dual usefulness; both for estimating the number of factors as part of estimator implementation but also for detecting remaining factors and cross-section dependence in the residuals $\hat u_{i,t}$.

\subsection{Strong and Weak Cross-section Dependence}
\cite{ChudikPesaranTosetti2011} define the strength of factor $f_{l,t}$ as a positive constant $0\leq\alpha_l\leq1$, named the exponent of cross-section dependence, which is determined by the loadings of the factor:\footnote{\cite{PesaranTosetti2011} relate a spatial process such as the spatial autoregressive (SAR) to a model with an infinite number of weak factors, see also Remark 3 in \cite{ChudikPesaranTosetti2011}. \cite{PesaranYang2020} connect strong cross-section dependence and dominant units.}
\begin{align}
    lim_{N\rightarrow\infty}N^{-\alpha_l}\sum_{i=1}^N|z_{i,l}| = K < \infty \label{eq:FactorStrength}
\end{align}
The factor is strong if $\alpha_l =1$, semi-strong for $1>\alpha_l>1/2$, semi-weak for $1/2>\alpha_l>0$ and weak for $\alpha_l=0$. An alternative interpretation of the degree of cross-section dependence is that, it is the logarithm of the share of non-zero factor loadings over number of cross-sections:
\begin{align}
 \alpha_l = \frac{ln\left(\sum_{i=1}^N\mathbf{1}(z_{i,l}\neq 0)\right)}{ln(N)},  
\end{align}
where $\mathbf{1}(z_{i,l}\neq 0)$ is an indicator function which takes the value one if the factor loading is non-zero \citep{Bailey2021}. In both cases, the overall strength of dependence in the panel is then determined by the strongest factor as:
\begin{align}
\alpha = max(\alpha_l).
\end{align}
If no factor structure is present in \eqref{eq:ie}, then $u_{i,t}$ is random noise and no cross-sectional dependence is left. Likewise, if weak dependence is present, then $u_{i,t}$ collapses to a simple additive fixed effects model. The occurrence of strong dependence implies a factor structure which goes beyond a simple TWFE model.\footnote{\cite{Kapetanios2023,Kapetanios2023a} discuss this finding from the perspective of the appropriate choice of an estimator, whether to use a robust TWFE or the CCEP estimator.}

\subsubsection{Estimation of the Exponent of Cross-Section Dependence}
Estimators for the exponent of cross-sectional dependence \(\alpha\) are proposed in \cite{BaileyKapetaniosPesaran2016} and \cite{BaileyKapetaniosPesaran2019}, both of which require the exponent to be \(1/2 < \alpha \leq 1\), implying that only the degree of (semi-) strong cross-section dependence can be estimated. \cite{BaileyKapetaniosPesaran2016} focus on the estimation of the exponent of a generic, observed variable, $x_{i,t}$. The estimate for $\hat{\alpha}$ is:
\begin{align}
    \hat{\alpha} = 1 + \frac{1}{2}\frac{ln(\hat{\sigma}^2_{\bar{x}})}{ln(N)}
\end{align}
where $\hat{\sigma}^2_{\bar{x}} = T^{-1}\sum_{t=1}^T\left(\bar{x}_t-\bar{x}\right)^2$, $\bar{x}_t = N^{-1}\sum_{i=1}^Nx_{i,t}$ and $\bar{x} = (NT)^{-1}\sum_{t=1}^T\sum_{i=1}^Nx_{i,t}$. The rate of convergence of the estimator is $ln(N)^{-1}$. The authors further propose a bias corrected version better suited for small samples as:
\begin{align}
    \dot{\alpha} = \hat{\alpha} - \frac{ln(\hat{\mu}_v^2)}{ln(N)}-\frac{\hat{\zeta}_N}{2[Nln(N)]\hat{\sigma}^2_{\bar{x}}}
\end{align}
with $\hat{\zeta}_N=N^{-1}\sum_{i=1}^N\hat{\sigma}_i^2$ and $\hat{\sigma}_i^2$ is the variance of the defactored by $m$ principal components $x_{i,t}$, where $m$ is the number of estimated factors. $\hat{\mu}_v^2$ is the average variance of cross-section averages that have significant non-zero loadings.

In a follow-up article, \cite{BaileyKapetaniosPesaran2019} propose a method to estimate the exponent of cross-sectional dependence of residuals. The estimator is a bit simpler:
\begin{align}
    \tilde{\alpha} &= \frac{ln(e_N'\tilde{\Delta}e_N)}{2 ln(N)},
\end{align}
where $e_N$ is a $N\times1$ vector of ones and $\tilde{\Delta} = (\tilde{\delta}_{ij})$ is a $N\times N$ matrix of significant pairwise correlations coefficients $\hat{\rho}_{ij}$ after multiple testing:
\begin{align}
    \tilde{\delta}_{ij} = \begin{cases}
    \tilde{\rho}_{ij}\text{, if } |\tilde{\rho}_{ij}| > cv(p,N)/\sqrt{T} \text{ and } i\neq j \\ 
    1 \text{, if } i = j \\ 
    0 \text{, otherwise}  \end{cases}
\end{align}
where $cv(sig,N)$ is a critical value from multiple testing in \cite{BaileyPesaranSmith2019} and $sig$ is the significance level. There is no closed form solution for the standard errors of $\tilde{\alpha}$. \cite{BaileyKapetaniosPesaran2019} propose a cross-sectional bootstrap to obtain standard errors. 

\subsubsection{Testing for Cross-Section Dependence}
Weak dependence implies a (TW)FE, while strong cross-sectional dependence a more complicated structure such as IFE, GFE or NSTW. As such, it is sufficient to estimate the strength category, (semi-) weak or (semi-) strong, while the actual strength is of less interest.  The CD-test of \cite{PesaranCD2015,Pesaran2021}  tests the null hypothesis of weak dependence against the alternative of strong cross-sectional dependence.\footnote{Formally the null hypothesis is \(0 \leq \alpha \leq (2-\epsilon)/4\), where \(\epsilon\) depends on the relative expansion rate of \(N\) and \(T\), \(T = O(N^\epsilon)\). For ``fixed'' \(T\) panels, the null boils down to \(0 \leq \alpha \leq 1/2\), while if \(N\) and \(T\) have the same rate, the null is \(0 \leq \alpha \leq 1/4\) \citep{PesaranCD2015}.} The test employs the sum of the pairwise residual $\hat u_{i,t}$ correlation coefficients  \(\rho_{i,j}\):
\begin{align}
CD = \sqrt{\frac{2T}{N(N-1)}} \sum_{i=1}^{N-1} \sum_{j=i+1}^N \rho_{i,j}.
\end{align}
Under the null hypothesis, the correlation coefficients are close to $0$ and the distribution of the test is asymptotically \(CD \sim N(0,1)\). If \(\alpha < 1/4\), the test has the correct size for all combinations of \(N\) and \(T\)  and it can be applied to models with lagged dependent variables. The test has good power for \(\alpha > 1/2\). However, it tends to over reject if \(\alpha\) is in the interval \((1/4,1/2]\) and if \(T\) is relatively large to \(N\). Notice that the $CD$ test is not only applicable to the residuals, but also directly to the observables themselves. If $y_{i,t}$ and $x_{i,t}$ are correlated with the unobserved heterogeneity, they will also be cross-sectionally dependent. 

\cite{JuodisReese2021} point out that if period-specific parameters such as fixed effects or cross-section averages are included in an estimation, the $CD$ test applied to the residuals diverges and over-rejects the null hypothesis. To regain asymptotic normality, they propose to reweigh the correlation coefficients with the Rademacher weights \(w_i\):
\begin{align}
\rho_{i,j} = \sum_{t=1}^T w_i \hat u_{i,t} \hat u_{j,t} w_j.
\end{align}		
It then follows that under the null hypothesis the test statistic is \(CD_w \stackrel{d}{\rightarrow} N(0,1)\). Since the weights are random, \cite{JuodisReese2021} propose to recalculate at least 30 times with varying weights the correlation coefficients, $\rho_{i,j}$, the $CDw$ statistic and in the last step use averaged weighted CD test as the final test statistic.
\cite{JuodisReese2021} propose a further extension based on the power enhancement method of \citep{FanLiaoYao2015} to overcome the potential under-powering for very large $N$:
\begin{align}
CD_{w+} = CD_w + \sum_{i=2}^N \sum_{j=1}^{i-1}|\hat{\rho}_{i,j}| \mathbf{1}\left(|\hat{\rho}_{i,j}| > 2 \sqrt{ln(N)T}\right).
\end{align}

As an alternative to solve the over-rejection issue of the $CD$ test applied to the residuals, \cite{Pesaran2021b} propose a bias-corrected version. The null hypothesis of the test is no cross-sectional dependence against the alternative of weak (or network) dependence in the presence of strong factors. The \(CD^*\) test is:
\begin{align}
CD^* = \frac{CD + \sqrt{\frac{T}{2}\theta}}{1-\theta}
\end{align}
The bias correction is dependent on \(\theta\) which is a function of the estimated factor loadings using principal components to approximate the factors.

\subsection{Direct Tests for Interactive Fixed Effects}
\cite{Bai2009} proposed a Hausman test of a TWFE against an IFE model, based on the difference between the TWFE and the ILS estimators:
\begin{align}
    J_B =  \left(\hat{\beta}_{TWFE} -\hat{\beta}_{ILS}\right)'V_B^{-1}\left(\hat{\beta}_{TWFE} -\hat{\beta}_{ILS}\right),
\end{align}
where $V_B = Var(\hat{\beta}_{ILS})-Var(\hat{\beta}_{TWFE})$. 
Under the null hypothesis of TWFE, the TWFE estimator is consistent and efficient, and $J_B\rightarrow_d \chi^2_k$.
As an alternative to the ILS estimator and in a fixed $T$ setting, \cite{Westerlund2019a} proposes using the CCEP estimator instead of the ILS estimator. The test statistic and the asymptotic distribution remain the same.

\subsection{Estimating $m$}
A key component in the implementation of many estimators, but also for detecting factors and cross-section dependence in the residuals, is the number of factors $m$. \cite{Bai2002} propose two information criteria for estimating $m$, which penalize the sum of squared residuals with a penalty \(g(N,T)\), which increases with each additional factor:
\begin{align}
PC(m) &= V(m,F^m) - m \hat{\sigma}^2  g(N,T), \\
IC(m) &= ln(V(m,F^m)) - m g(N,T),\\
\intertext{where there are three choices for \(g(N,T)\):}
g(N,T) &=  \begin{cases}
\frac{N+T}{NT} ln\left(\frac{NT}{N+T}\right),\\
\frac{N+T}{NT} ln\left(\min(N,T)\right),\\
\frac{\ln\left(\min(N,T)\right)}{\min(N,T)}.\\
\end{cases}
\end{align}
In the above expression, \(V(m,\hat{F}^m) = \frac{1}{N} \sum_{i=1}^N \sum_{t=1}^T \hat{e}_{i,t}^2\) and \(\hat{e}_{i,t}\) are residuals of the variable of interest on the first \(m\) first principal components. The criteria are calculated for \(m=1,...,m_{max}\) and $\hat m$ is the number of factors which minimizes the criteria. The method is known to overestimate the true $m$ \citep{Juodis2022a}. 

\cite{AhnHorenstein2013} point out two disadvantages in IC and PC. First, the penalty functions are pre-specified and not data dependent \citep{HallinLiska2007, Onatski2010}. Second, the maximum number of common factors is chosen by the researcher in an ad hoc way. To overcome the issues, they suggest the Eigenvalue Ratio (\(ER(k)\)) and Growth Rate (\(GR(k)\)) estimators. Both take the ratio of residual variances when an additional common factor is added into account. The \(ER(k)\) uses the \(k\)th largest eigenvalues of \((NT)^{-1}\sum_{i=1}^NX_i'X_i\) while the \(GR(k)\) is based on the mean of the squared residuals of a regression of \(X_i\) on the first \(k\) principal components of \((NT)^{-1}\sum_{i=1}^NX_i'X_i\). The number of factors is selected as:
\begin{align}
\tilde{m}_{ER} &= max_{1\leq m \leq m_{max}} ER(m),\\ \tilde{m}_{GR} &= max_{1\leq m \leq m_{max}} GR(m) 
\end{align}

A third method is proposed by \cite{Onatski2010}. Here the idea is to take the differences between the ordered eigenvalues of the variable of interest into account. The number of factors is estimated as the first difference which is greater than a threshold. The advantage of the method is that it allows weak dependence across both time and the cross-section, and also for integrated factors.

Finally, a fourth method would be applying the adaptive group LASSO of \cite{LuSu2016}. Their Monte Carlo simulations show that for $N =T= 20$ the LASSO method correctly identifies the true number of factors 54\% of the time. In comparison, \cite{Bai2002} criteria have correct rates between 0\% and 14\%, Onatski's (2010) method achieves 22\%, and the ER and GR estimators from \cite{AhnHorenstein2013} have correct rates of 23\% and 25\%, respectively.

An estimator for the number of common factors in residuals is proposed by \cite{Gagliardini2019}. In the first step, the method checks if there is a common factor structure, whose number it estimates in the second step. In both steps, the difference between the largest eigenvalue, \(\mu(m)\), and the penalty, \(g(\cdot,\cdot)\), is used:
\begin{align}
	\xi(m) &= \mu(m) - g(N,T), \\
	g(N,T) &= \frac{\left(\sqrt{N}+\sqrt{T}\right)^2}{NT} ln\left( \frac{NT}{\left(\sqrt{N}+\sqrt{T}\right)^2}\right).
\end{align}
The number of factors is then estimated as:
\begin{align}
	\text{no factors}&: \xi(1)<0 \\
	\text{\(\hat{k}\) factors}&: \hat{m}_{GOS} = min(0,...,T-1:\xi(k)<0)
\end{align}
While the methods by \cite{Bai2002,AhnHorenstein2013,Onatski2010} are flexible on the relative rates of \(N\) and \(T\), \cite{Gagliardini2019} requires \(N>T\). 

\subsection{Slope Heterogeneity}
In the same way that long periods of time introduce price variations to the time-invariant individual effects, it is also reasonable to assume time variation in the coefficients. And just as there is individual heterogeneity across units, it is also possible to have heterogeneity in the slopes. The literature on these topics is too large to cover here, so we briefly present some heterogeneity tests, especially for the IFE model, which is the oldest and most developed of the three. Many of the estimators presented in Section \ref{sec:estim} can be adapted for this case. 

Cross-section slope heterogeneity implies $\beta_{i}\neq \beta_{j}$ for units $i$ and $j$. If the time dimension is sufficiently large, the $\beta_i$ can be individually estimated. Tests for slope heterogeneity test the null hypothesis \(\beta_i=\beta\) against the alternative that \(\beta_i\neq\beta_j\) for some $i\neq j$.\footnote{\cite{bonhommedenis2025} give a comprehensive overview over the topic. \cite{Fernández-ValLee2013} discuss estimation of the mean and variance of $\beta_i$ in a model without interactive effects.} The literature on such tests is scarce. In the presence of IFE, \cite{SuChen2013} develop a residual-based LM test in which the restricted, homogeneous-coefficient model, is estimated with the ILS estimator. Another contribution is that by \cite{PesaranYamagata2008}, but for the TWFE model. The test takes the difference between the individual slopes and the pooled estimate into account and requires large $N$ and large $T$. \cite{Blomquist2013} extend the latter test to allow for heterosekdasticity and autocorrelation. \cite{BersvendsenDitzen2021} provide Monte Carlo evidence that it can also be applied to an IFE setting. Several of the methods presented above can be adapted to estimate heterogeneous coefficients, such as CCE and TSIV. GFE can be employed when coefficients vary across groups, in the presence of GFE. An alternative methodology is developed in \cite{SuShiPhillips2016}, based on a classifier Lasso which shrinks individual coefficients to the group-specific coefficients, but in the case of FE.   

The time series equivalent for heterogeneous slopes in space, is parameter instability. While there are many types of parameter instability, the most popular in economics is that of abrupt structural breaks. A structural break implies that the causal relationship between the explanatory variable(s) and the dependent variable changes at discrete points in time. Ignoring structural breaks can lead to model misspecification and biased estimation.  Structural changes or ``breaks'' are widespread in macroeconomic and financial time series \citep{StockWatson1996}. These could occur as the outcome of technological or legislative changes, policy changes, macroeconomic shocks, or natural disasters and wars, and can be observed, subtle and detected later or entirely unobserved.\footnote{Structural changes can also be related to the Difference in Differences literature, see for example \cite{ROTH20232218}.} 

The number and exact location of the structural breaks can be estimated and tested. \cite{NKW2022} and \cite{DKW2025} propose an estimator for the number and location of breaks, confidence intervals, and tests for detecting the existence of breaks. A variant of the CCEP estimator is used, based on the IFE model. \cite{Li2016} and \cite{Kaddoura2022} show how structural breaks can be estimated using the Lasso, and  \cite{SarafidisZhuSilvapulle2020} offer a test which works for fixed $T$. 

Mixing structural breaks and individual slopes is also possible. \cite{BALTAGI201787,Baltagi2021} show how to estimate the break date in the presence of parameter heterogeneity, using the CCE estimator.\footnote{\cite{OKUI2021447,LUMSDAINE202345} provide methods for estimating structural breaks for models with group-specific slope coefficients but no interactive fixed effects.}   
Another form of time heterogeneity popular in economic models is that of regime switching. The business cycles in economics and the boom and bust cycles in finance are examples where it is reasonable to approximate the state of the economy by different regimes. This type of modeling is also widespread in economics. 

Threshold regression in panel data with interactive fixed effects takes the form:
\begin{align}
	y_{i,t} &=  \mathbf{x}'_{i,t} \boldsymbol{\beta}_{1} + u_{i,t}, \; \text{if}\; w_{i,t}\leq \tau,\\
	y_{i,t} &=  \mathbf{x}'_{i,t} \boldsymbol{\beta}_{2} + u_{i,t}, \; \text{if}\; w_{i,t}\geq \tau,
\end{align}
where $w_{i,t}$ is an observed variable, potentially one of $x_{i,t}$, and $\tau$ is the threshold parameter, which is estimable from the data. In the model above the coefficients switch according to whether the variable $ w_{i,t} $ is above or below a threshold. \cite{ChudikEtAl2017} and \cite{Miao2020} show how to estimate the above model using the CCEP and ILS estimators respectively and provide tests of nonlinearity.\footnote{\cite{Miao2020} consider a threshold regression in panels with group structure but not IFE.} \cite{DKW2025threshold} extend the CCEP estimator to homogeneous threshold models while \cite{BarassiKaraviasZhu2025} extend the above model to heterogeneous coefficients and heterogeneous thresholds. 

\section{Two Empirical Investigations}
This section investigates two important research questions using the methods presented earlier. Our aim is two-fold; first, to examine if there is evidence of the general forms of unobserved heterogeneity presented in this paper and how the estimation results based on the new estimators compare to the results of FE and TWFE. Second, to demonstrate that there is available software already available. A list of software commands is presented in the Appendix at the end.

The analysis for both applications includes the following estimators. For traditional one-way and two-way error component models we use the FE and TWFE estimators. For IFE we employ ILS with $1-3$ factors for robustness, CCEP and TSIV. For GFE we use the GF estimator, and finally, for NSTW we use the ILS with $8-10$ factors and TSGF-KT. In terms of model diagnostics, we employ the CD, CDw and $CD^*$ tests of cross-section dependence, the ER, GR and GOS information criteria for selecting the number of factors. Finally, we also test explicitly for the rank condition in CCEP.\footnote{The analysis took place in Stata using the software presented in the appendix.}  

\subsection{Inflation, Growth, and the Shadows Between}
The post-pandemic high inflation  regime brought into the spotlight again the relationship between inflation and economic growth. This issue is pressing as economic growth has been consistently below the trend after the global financial crisis, and shows no signs of recovery, in turn increasing poverty, inequality, and leading governments to raise taxes in order to fund public services like education, pensions, healthcare and defense. 

The relationship between inflation and economic growth has been the subject of much research, including \cite{fischer1993}, \cite{degregorio1993} and \cite{barro1995}, inter alia. Initially, the relationship was found to be negative, a result which was later overturned in favor of the existence of a nonlinear relationship, where inflation below a certain threshold has a positive effect on growth, while inflation above that threshold has a negative effect. There have been many attempts to estimate this threshold, with one of the most cited ones being \cite{khansenhadji2001}, which was among the first to apply \cite{hansen1999} threshold regression approach in the present context. Using panel data covering $140$ countries between $1960$ and $1998$ they estimate the threshold to $11$\% for the full sample or for developing economies and to $1$\% for industrialized countries.

In this paper we estimate the growth-inflation nexus assuming FE, TWFE, IFE, GFE and NSTW in the error term $u_{i,t}$. The regression model follows \cite{khansenhadji2001}:
\begin{align}\label{app:inflmod}
	\Delta GDP_{i,t}&=INFL^*_{i,t}\mathbb{I}\left(INFL^*_{i,t}\leq \tau\right)+INFL^*_{i,t}\mathbb{I}\left(INFL^*_{i,t}>\tau\right)+\beta_3 O_{i,t}\\
    &+\beta_4 \Delta POP_{i,t}+\beta_5 CAP_{i,t}+\beta_6 GOV_{i,t}+u_{i,t},
\end{align}
and is estimated using the dataset from \cite{DKW2025threshold} which includes $74$ countries observed over the years $1970-2022$,  collected from the World Bank Development Indicators database.\footnote{For robustness checks, the sample is further split into two subsamples of $51$ non-industrialized and $23$ industrialized countries. The main findings remain the same and as such are not presented, but they are available upon request.} The dependent variable is GDP per capita growth ($\Delta GDP$), while the independent variables are trade openness ($O$), population growth ($\Delta POP$), gross capital formation ($CAP$) and government spending ($GOV$). The regressor of interest is inflation ($INFL^*$) above and below the threshold of 1 defined as: 
\begin{align}
	INFL_{i,t}^*=(INFL_{i,t}-1)\mathbb{I}(INFL_{i,t}\leq 1)+log(INFL_{i,t})\mathbb{I}(INFL_{i,t}> 1),	
\end{align}
where $INFL$ is inflation. The definition of $INFL^*$ is based on a partial log transformation, where the logarithmic transformation is applied on $INFL$, but only for values greater than $1$. \cite{ghoshphillips1998} argue that the logarithmic transformation provides the best fit in a class of non-linear models.

The threshold parameter $\tau$ is unknown, and thus we estimate it in a first step using the CCEP-based threshold regression methodology of \cite{DKW2025xtthres}. The estimated threshold for the full sample is $9.39$\% which is in line with the results of \cite{khansenhadji2001}. The threhsold parameter estimator is shown to be superconsistent in \cite{DKW2025xtthres} and thus we will henceforth assume it is known in the subsequent analysis.

The $CD$ test in Table \ref{table:csd_inf} below strongly rejects the null hypothesis of cross-section independence for all variables. The $ER$ and $GR$ estimate $1$ factor in all variables except Capital, where the estimated number of factors is equal to the maximum number of factors permitted ($8$).

Table \ref{table:inf_main} presents the results for the 12 estimators. Overall, there is broad estimator agreement to the signs of the coefficients, with high inflation, population growth and government spending having a negative effect on growth, while capital formation and trade openness having a positive one. These signs are in line with economic theory and previous studies. The sign of inflation oscillates across estimators for the low inflation regime and in all but the FE estimator case, there is no statistical significance. Despite the sign consistency, the coefficient differences across estimation methods are striking. The inflation coefficient in the high regime for example, is equal to $-0.210$ for the FE, almost doubles to $-0.377$ for the TWFE and then almost doubles again for the TSGF-KT to $-0.607$. The ILS estimators with 1, 2, 3, 8, and 9 factors hover around $-0.43$, while the ILS(10) is at $-0.527$, closer to the TSGF-KT. Similar values are obtained by the CCEP ($-0.567$) and smaller in absolute value for the TSIV ($-0.479$). The CCEP being closer to TSGF-KT when compared to ILS could be due to its smaller bias under NSTW as can be seen from the Monte Carlo experiments in \cite{Beyhum2024}.\footnote{We used $1,000$ initial values for the implementation of TSGF-KT. While the estimates are the outcome of convergence to a local minimum, the estimator differences across these minima were small and economically insignificant. Table 2 presented the estimate with the minimum SSR over these $1000$ estimates. The initialization happened by selecting different seeds and not, for example, by searching in the parameter space.} This discrepancy begs for further analysis in terms of the model specification.

Looking at regression residuals, the $CD$ test strongly rejects the null hypothesis of no cross-section dependence at the $5$\% level for all estimators except for the ILS(8)-ILS(10), and the TSGF-KT. Interestingly, all estimators are for NSTW models. We complement this line of enquiry on the residuals with the $CD^*$ test by \cite{Pesaran2021b}, and additionally by estimating the number of factors in the residuals with the $ER$ and $GOS$ statistics. Both methods find residual cross-section dependence in almost all cases. Notable exceptions are the ILS(1)-ILS(3) and TSGF-KT for which $CD$ and $CD^*$ do not reject. The number of factors in the residuals estimated by $ER$ and $GOS$ is $0$ for all estimations with the exception of GF. Some further comments are in order. The CCEP estimator passes the rank condition test. The ILS estimator requires more iterations as the number of factors increases. The discretization split the countries into $32$ groups and the time periods into $2$ groups. The GF estimator, coupled with the BIC criterion, splits the sample into four groups. The main countries in the first group are the US, Canada, South Africa, Australia, Japan and most of the EU, the second group includes Mexico, India, Pakistan, Indonesia, Egypt and some countries in south America. The third group contains Saudi Arabia and Gabon only, while the fourth group is a singleton containing only Rwanda. 

Overall, there is strong evidence that high inflation has a negative effect on growth and it seems that NSTW is the most appropriate model for capturing unobserved heterogeneity.\footnote{We further split the sample into developed and developing economies and performed the same analysis. The main conclusions are similar and hence the results are omitted from the paper, but are available upon request.}

\begin{table}[!h]
    \centering
    \begin{tabular}{l cc cc }\hline\hline
    \toprule
    Variable & \multicolumn{2}{c}{Test for Cross-Sectional Dependence}& \multicolumn{2}{c}{Number of factors}\\
     & CD & p & ER & GR \\ \cmidrule(lr){2-3} \cmidrule(lr){4-5}
    GDP & 72.87 & 0.00 & 1 & 1 \\
Inflation & 127.03 & 0.00 & 1 & 1 \\
Openess & 86.47 & 0.00 & 1 & 1 \\
Pop Growth & 71.70 & 0.00 & 1 & 1 \\
Capital & 15.21 & 0.00 & 8 & 8 \\
Gov. Spend & 37.01 & 0.00 & 1 & 1 \\
All & & & 1 & 1 \\
\hline\hline

    \end{tabular}
    \caption{CD Test Statistic and Number of Factors. CD based on \cite{PesaranCD2015}, ER stands for Eigenvalue Ratio and GR for Growth Ratio from \cite{AhnHorenstein2013}.}  
    \label{table:csd_inf}
\end{table}

\begin{sidewaystable}
 \centering
    \resizebox{\textwidth}{!}{%
    \centering
    \begin{tabular}{l c cc ccc cc cc cc cc cc cc cc }\hline\hline
    \toprule
     &(1)   &(2)   &(3)   &  (4)   & (5)   & (6)   & (7)   & (8)   & (9)   & (10) & (11) & (12)  \\  
    & FE & TWFE & ILS(1) & ILS(2) & ILS(3) & ILS(8) & ILS(9) & ILS(10)& CCEP & TSIV &  TSGF-KT & GF \\  
    \cmidrule(lr){2-11} \cmidrule(lr){12-13}
    Inflation Low    &       0.348   &       0.111   &       0.077   &       0.021   &      -0.018   &      -0.101   &      -0.089   &      -0.095   &       0.116   &       0.019   &      -0.053   &       0.119   \\
            &     (0.071)***&     (0.071)   &     (0.069)   &     (0.068)   &     (0.066)   &     (0.067)   &     (0.067)   &     (0.068)   &     (0.102)   &     (0.081)   &     (0.068)   &     (0.069)   \\
Inflation High   &      -0.210   &      -0.377   &      -0.402   &      -0.419   &      -0.465   &      -0.465   &      -0.469   &      -0.527   &      -0.567   &      -0.479   &      -0.607   &      -0.352   \\
            &     (0.067)** &     (0.076)***&     (0.069)***&     (0.067)***&     (0.065)***&     (0.069)***&     (0.069)***&     (0.070)***&     (0.103)***&     (0.095)***&     (0.074)***&     (0.067)***\\
Openness        &       0.003   &       0.009   &       0.014   &       0.011   &       0.016   &       0.033   &       0.047   &       0.047   &       0.025   &       0.037   &       0.033   &       0.004   \\
            &     (0.003)   &     (0.003)** &     (0.003)***&     (0.003)***&     (0.003)***&     (0.004)***&     (0.004)***&     (0.004)***&     (0.009)** &     (0.011)***&     (0.005)***&     (0.001)***\\
Population Growth       &      -0.434   &      -0.589   &      -0.726   &      -0.673   &      -0.930   &      -0.985   &      -0.926   &      -0.903   &      -0.827   &      -0.715   &      -0.778   &      -0.672   \\
            &     (0.081)***&     (0.080)***&     (0.081)***&     (0.078)***&     (0.097)***&     (0.098)***&     (0.100)***&     (0.101)***&     (0.311)** &     (0.194)***&     (0.076)***&     (0.071)***\\
Gross Capital    &       0.132   &       0.145   &       0.150   &       0.167   &       0.162   &       0.120   &       0.113   &       0.104   &       0.102   &       0.109   &       0.140   &       0.139   \\
            &     (0.015)***&     (0.014)***&     (0.014)***&     (0.014)***&     (0.013)***&     (0.015)***&     (0.015)***&     (0.016)***&     (0.029)***&     (0.035)** &     (0.014)***&     (0.014)***\\
Gov. Spending    &      -0.273   &      -0.217   &      -0.181   &      -0.172   &      -0.182   &      -0.236   &      -0.285   &      -0.298   &      -0.302   &      -0.248   &      -0.165   &      -0.085   \\
            &     (0.024)***&     (0.023)***&     (0.024)***&     (0.023)***&     (0.023)***&     (0.026)***&     (0.028)***&     (0.028)***&     (0.057)***&     (0.070)***&     (0.027)***&     (0.014)***\\
\hline Obs    &        3922   &        3922   &        3922   &        3922   &        3922   &        3922   &        3922   &        3922   &        3922   &        3922   &        3922   &        3922   \\
N      &          74   &          74   &          74   &          74   &          74   &          74   &          74   &          74   &          74   &          74   &          74   &          74   \\
T           &          53   &          53   &          53   &          53   &          53   &          53   &          53   &          53   &          53   &          53   &          53   &          53   \\
\hline CD   &      69.410   &      -2.034   &      10.334   &       4.139   &       3.074   &      -1.293   &      -1.350   &      -1.744   &      65.767   &      84.805   &      -0.029   &     195.956   \\
\ \ p       &       0.000   &       0.042   &       0.000   &       0.000   &       0.002   &       0.196   &       0.177   &       0.081   &       0.000   &       0.000   &       0.977   &       0.000   \\
CDw         &      -0.109   &      -0.997   &      -0.610   &      -1.080   &      -0.904   &      -0.787   &       0.159   &       0.975   &      -2.159   &       0.791   &      -0.492   &       0.113   \\
\ \ p       &       0.914   &       0.319   &       0.542   &       0.280   &       0.366   &       0.431   &       0.873   &       0.330   &       0.031   &       0.429   &       0.623   &       0.910   \\
CD*         &       3.976   &      -3.529   &      -0.765   &      -1.596   &      -0.639   &      -2.628   &      -2.724   &      -2.852   &       2.842   &       3.913   &       1.305   &       0.197   \\
\ \ p       &       0.000   &       0.000   &       0.444   &       0.111   &       0.523   &       0.009   &       0.006   &       0.004   &       0.004   &       0.000   &       0.192   &       0.844   \\
OIR         &               &               &               &               &               &               &               &               &               &           0   &               &               \\
\ \ p       &               &               &               &               &               &               &               &               &               &           -   &               &               \\
$\hat{m}_{ER}$&           0   &           0   &           0   &           0   &           0   &           0   &           0   &           0   &           0   &           0   &           0   &           1   \\
$\hat{m}_{GOS}$&           0   &           0   &           0   &           0   &           0   &           0   &           0   &           0   &           0   &           0   &           0   &           1   \\
RC          &               &               &               &               &               &               &               &               &           1   &               &               &               \\
Clusters (units)&               &               &               &               &               &               &               &               &               &               &           32   &           4   \\
Clusters (time)&               &               &               &               &               &               &               &               &               &               &           2   &               \\
Converged   &           1   &           1   &        true   &        true   &        true   &        true   &        true   &        true   &               &               &               &               \\
Iter.       &               &               &          32   &          34   &          56   &         177   &         181   &         251   &               &               &               &               \\
    CSD & No & No & $PC(1)$ &  $PC(2)$ &  $PC(3)$ &$PC(8)$ &  $PC(9)$ &  $PC(10)$ & $\mathbf{\overline{Z}}_t$ & IV &  k-means-TWFE & k-means \\  \hline\hline
    \end{tabular}}
    \caption{Notes. Significance levels: \(^* \%5, ^{**}1\%, ^{***}0.01\%\). $CD$ test for weak cross-sectional dependence of residuals. $RC$ denotes the rank condition classifier in \citep{devossarafidis2024}. $\widehat{m}$ denotes the estimated number of factors in $\hat u_{i,t}$ using the Eigenvalue Ration ($ER$) criterion from \cite{AhnHorenstein2013} and the criterion from \cite{Gagliardini2019} ($GOS$). CSD denotes how cross-section dependence is addressed. $PC(j)$ stands for principal components \citep{Bai2009} with $j$ factors, $[\bar y, \bar X]$ for cross-section averages \citep{Pesaran2006}, and IV for instruments \citep{NorkuteEtal2021}. k-means-TWFE for the TSGF-KT estimator of \cite{Beyhum2024}, and k-means for GF \citep{Bonhomme2015a}. Threshold for inflation is 9.$39$\%. } \label{table:inf_main} 
\end{sidewaystable}

\subsection{Capital Mobility Reassessed: New Evidence on the Feldstein-Horioka Puzzle}
In this section we study the Feldstein-Horioka puzzle \citep{feldsteinhorioka1980}, which is one of the six main puzzles in international macroeconomics according to \cite{obstfeldrogoff2000}. \cite{feldsteinhorioka1980} presented statistical evidence that the correlation between a country's domestic savings and its domestic investment is high, which would be a puzzle in a world of high capital mobility. Instead, the expectation is that savings flow across borders to wherever returns are highest. At the same time, a country's investment level should have been uncoupled from its domestic savings as it would be able to finance investment through foreign capital. 

\cite{feldsteinhorioka1980} provided as evidence for the puzzle the slope coefficient of savings from the regression of investment on savings. They also considered the impact of trade openness and potential nonlinearity in the relationship, induced by high levels of trade openness. Here we will employ a variant of the original \cite{feldsteinhorioka1980} model which allows for unobserved heterogeneity in the forms of IFE, GFE and NSTW unobserved heterogeneity models:
\begin{align}\label{app:mod}
I_{i,t}&=\beta_{1}S_{i,t}+\beta_{2}O_{i,t}+\beta_{3}O_{i,t}S_{i,t}+u_{i,t},
\end{align}
where $I_{i,t}$ is domestic investment, $S_{i,t}$ is domestic savings, and $O_{i,t}$ is trade openness (sum of imports and exports), all as percentage of GDP. \cite{feldsteinhorioka1980} do not include $O_{i,t}S_{i,t}$ as a regressor outside the interaction term $O_{i,t}S_{i,t}$, but we include it here to avoid omitted variable bias. The data are collected from the World Bank's World Development Indicators for a balanced panel of $98$ countries observed over the $40$ years $1980-2019$. 

Table \ref{table:csd_fh} presents the CD test of \cite{PesaranCD2015} for all the variables in the above model. The test rejects at the $1$\% level for all, showing evidence of cross-section dependence. The $ER$ and $GR$ statistics from \cite{AhnHorenstein2013} are employed to detect the number of factors in each variable. The two ratios agree that there are two unobserved factors driving investment and one factor driving trade openness. However, when it comes to savings, the $ER$ statistic finds one factor while the $GR$ statistic finds $5$ factors. The interaction term however still has one factor according to both ratios, hence we would think that the $ER$ statistic's results on the factors in savings are more reliable. Having two factors in the dependent variable and one in the independent  would also mean that the rank condition of the CCE estimators is satisfied.

Table \ref{table:fh_main} presents the estimation results for the model \eqref{app:mod}. The signs are consistent across estimators and in line with \cite{feldsteinhorioka1980} and the subsequent literature. For all estimators with the slopes of $S_{i,t}$ and $O_{i,t}$ are positive and the slope of $S_{i,t} O_{i,t}$ is negative. 

Observing the estimated values of the savings rate (the savings coefficient), we once more observe a wide range, from $0.155$ for the $ILS(10)$ estimator, which allows for NSTW, but is inefficient, up to $0.437$ for the TSGF-KT. These values are much lower than the $0.89$ found in \cite{feldsteinhorioka1980} and $0.6$ from \cite{obstfeldrogoff2000}. However, they are in line with \cite{wacziargvamvakidis1998} who estimate it at $0.423$ for a sample of $103$ countries between $1970$ and $1993$, and close to $0.477$ of \cite{lusu2023}, based on an unbalanced sample of $135$ countries observed from $1975$ to $2017$. The latter allows for slope coefficient heterogeneity across units and time, but not for IFE/GFE/NSTW. 

Discrepancies across estimators appear here as well. For example, the TWFE estimate of the trade openness's coefficient is $0.07$, while the NSTW-KT is almost double at $0.131$. The regression diagnostics indicate remaining cross-section dependence in most residuals. Interestingly, all $CD$-type tests, and the number of factor estimators, agree that there is no remaining cross-section dependence in the ILS(10) and TSGF-KT residuals, where NSTW heterogeneity has been controlled for. This is evidence that the error structure is more complicated than the IFE model, potentially by the existence of many weak factors. Overall, the above results suggest that the NSTW model is more suitable as it is capable of removing the unobserved heterogeneity. The ILS(9) and ILS(10) coefficients are smaller than that of TSGF-KT, but the latter is trustworthy given that it is a more efficient estimator. 

Table \ref{table:fh_main} presents some additional results. The rank condition in the CCE estimator is again satisfied. The number of iterations until convergence for the ILS estimators increases with the number of factors but that increase did not result in noticeably worse computational time, at least for this sample which is relatively small.

In the process of approximating the non-separable heterogeneity by a latent group structure the discretization of the heterogeneity in TSGF-KT finds 15 groups, which means that there is significant heterogeneity among the $98$ countries. A second finding from the discretization is that there are two time clusters, hinting at reduced variation across time. The discretization results find a time effect before $1995$ and one after that. To examine if there is a structural break at the coefficients as well, we test for breaks using the methods of \cite{DKW2025}. Table \ref{table:fh_breaks} performs the $UDmax$ test of the null of no breaks against the alternative of up to $5$ breaks in the sample. The test is performed in the whole coefficient vector to detect any instabilities, rejecting the null of no breaks. The sequential test statistic reveals a single break in 1996 with a $95$\% confidence interval $[1995,1997]$. This is evidence that a break in the factors occurred at the same time as a break in the coefficients. 

Table \ref{table:fh_breaks_estim} estimates the model again assuming a break in $1996$, with the CCEP estimator. It is interesting to note that there is a downward movement of the savings slope coefficient after $1996$, showing that the Feldstein-Horioka puzzle does not persist in time. Furthermore, the pre-break coefficient is at $0.454$, higher than in the results from Table \ref{table:fh_main}, but still low regarding most of the literature. Interestingly, it seems that the break affected only the savings coefficient and not the trade openness ones.\footnote{We have rerun the regressions in Table \ref{table:fh_main} for the developed and developing countries and these results are available upon request.} 

\begin{table}[]
    \centering
    \begin{tabular}{c cc cc c}\hline\hline
    & C & S & O  & S $\times$ O & All \\\hline
    \multicolumn{6}{l}{Test for Cross-Sectional Dependence}\\ \hline
    CD & 21.02 & 8.52 & 85.52 & 47.63 & \\
p & 0.00 & 0.00 & 0.00 & 0.00 & \\
\hline \multicolumn{6}{l}{Number of Factors}\\\hline ER & 2 & 1 & 1 & 1 & 1 \\
GR & 2 & 5 & 1 & 1 & 1 \\
\hline\hline

    \end{tabular}
    \caption{$CD$ Test Statistic and Number of Factors. $CD$ based on \cite{PesaranCD2015}, $ER$ stands for Eigenvalue Ratio and $GR$ for Growth Ratio from \cite{AhnHorenstein2013}. \textit{C} is Capital, \textit{S} Savings, \textit{O} Openness.}  
    \label{table:csd_fh}
\end{table}

\begin{sidewaystable}
 \centering
    \resizebox{\textwidth}{!}{%
    \centering
    \begin{tabular}{l c cc ccc cc cc cc cc cc cc }\hline\hline
    \toprule
        &(1)   &(2)   &(3)   &  (4)   & (5)   & (6)   & (7)   & (8)   & (9)   & (10) & (11) & (12)  \\  
    & FE & TWFE & ILS(1) & ILS(2) & ILS(3) & ILS(8) & ILS(9) & ILS(10)& CCEP & TSIV &  TSGF-KT & GF \\  
    \cmidrule(lr){2-11} \cmidrule(lr){12-13}
    Savings     &       0.400   &       0.397   &       0.355   &       0.342   &       0.277   &       0.212   &       0.237   &       0.155   &       0.385   &       0.378   &       0.437   &       0.416   \\
            &     (0.017)***&     (0.017)***&     (0.017)***&     (0.016)***&     (0.017)***&     (0.019)***&     (0.019)***&     (0.019)***&     (0.154)*  &     (0.114)***&     (0.019)***&     (0.016)***\\
Openness    &       0.074   &       0.070   &       0.113   &       0.123   &       0.120   &       0.074   &       0.091   &       0.068   &       0.091   &       0.127   &       0.131   &       0.065   \\
            &     (0.005)***&     (0.006)***&     (0.006)***&     (0.007)***&     (0.007)***&     (0.008)***&     (0.008)***&     (0.008)***&     (0.023)***&     (0.027)***&     (0.008)***&     (0.004)***\\
OpenSave    &      -0.002   &      -0.002   &      -0.002   &      -0.002   &      -0.002   &      -0.001   &      -0.001   &      -0.001   &      -0.002   &      -0.003   &      -0.003   &      -0.002   \\
            &     (0.000)***&     (0.000)***&     (0.000)***&     (0.000)***&     (0.000)***&     (0.000)***&     (0.000)***&     (0.000)***&     (0.001)*  &     (0.001)*  &     (0.000)***&     (0.000)***\\
\hline Obs    &        3920   &        3920   &        3920   &        3920   &        3920   &        3920   &        3920   &        3920   &        3920   &        3920   &        3920   &        3920   \\
N      &          98   &          98   &          98   &          98   &          98   &          98   &          98   &          98   &          98   &          98   &          98   &          98   \\
T           &          40   &          40   &          40   &          40   &          40   &          40   &          40   &          40   &          40   &          40   &          40   &          40   \\
\hline CD   &      23.316   &       0.771   &      30.152   &      11.320   &       7.991   &       5.661   &      10.973   &      -0.710   &      -1.493   &      23.001   &      -1.409   &     176.559   \\
\ \ p       &       0.000   &       0.441   &       0.000   &       0.000   &       0.000   &       0.000   &       0.000   &       0.478   &       0.135   &       0.000   &       0.159   &       0.000   \\
CDw         &      -2.229   &      -0.364   &      -0.826   &      -1.544   &       3.199   &      -2.490   &       3.349   &       0.323   &      -0.235   &       4.468   &       0.165   &      -3.227   \\
\ \ p       &       0.026   &       0.716   &       0.409   &       0.123   &       0.001   &       0.013   &       0.001   &       0.747   &       0.814   &       0.000   &       0.869   &       0.001   \\
CD*         &      10.669   &      -2.274   &      10.692   &       2.084   &       2.295   &       0.687   &      -0.972   &       1.618   &      -0.094   &       8.063   &      -0.823   &      10.046   \\
\ \ p       &       0.000   &       0.023   &       0.000   &       0.037   &       0.022   &       0.492   &       0.331   &       0.106   &       0.925   &       0.000   &       0.411   &       0.000   \\
$\hat{m}_{ER}$&           1   &           1   &           6   &           0   &           0   &           0   &           0   &           0   &           2   &           1   &           0   &           3   \\
$\hat{m}_{GOS}$&           1   &           1   &           1   &           1   &           0   &           0   &           0   &           0   &           1   &           1   &           0   &           2   \\
RC          &               &               &               &               &               &               &               &               &           1   &               &               &               \\
Clusters (units)&               &               &               &               &               &               &               &               &               &               &           15   &           7   \\
Clusters (time)&               &               &               &               &               &               &               &               &               &               &           2   &               \\
Converged   &           true   &           true   &        true   &        true   &        true   &        true   &        true   &        true   &               &               &               &               \\
Iterations     &               &               &          38   &          56   &          75   &         292   &         224   &         439   &               &               &               &               \\ %%
    CSD & No & No & $PC(1)$ &  $PC(2)$ &  $PC(3)$ &$PC(8)$ &  $PC(9)$ &  $PC(10)$ & $\mathbf{\overline{Z}}_t$ & IV &  G-TWFE &GFE \\  \hline\hline
    \end{tabular}}
    \caption{Notes. Significance levels: \(^* \%5, ^{**}1\%, ^{***}0.01\%\). $CD$ test for weak cross-sectional dependence of residuals. $RC$ denotes the rank condition classifier in \citep{devossarafidis2024}. $\widehat{m}$ denotes the estimated number of factors in $\hat u_{i,t}$ using the Eigenvalue Ration ($ER$) criterion from \cite{AhnHorenstein2013} and the criterion from \cite{Gagliardini2019} ($GOS$). CSD denotes how cross-section dependence is addressed. $PC(j)$ stands for principal components \citep{Bai2009} with $j$ factors, $[\bar y, \bar X]$ for cross-section averages \citep{Pesaran2006}, and IV for instruments \citep{NorkuteEtal2021}. k-means-TWFE for the TSGF-KT estimator of \cite{Beyhum2024}, and k-means for GF \citep{Bonhomme2015a}. }  
    \label{table:fh_main}
\end{sidewaystable}

\begin{table}[]
    \centering
    \begin{tabular}{c ccc}\hline\hline
    & \multicolumn{3}{c}{Estimated Breaks}\\
    \cmidrule(lr){2-4}
   UDmax & Breakpoint&\multicolumn{2}{c}{$95\% CI$}\\
   \cmidrule(lr){1-1} \cmidrule(lr){2-2} \cmidrule(lr){3-4}
         7.776 & 1996 & 1995 & 1997 \\
\hline\hline

    \end{tabular}
    \caption{UDmax: Test of no breaks against the alternative of up to 5 breaks. Estimated break point and its $95\%$ confidence interval. Estimated equation: $C_{i,t} = \beta_1 S_{i,t} + \beta_2 O_{i,t} + \beta_3 S_{i,t}\times O_{i,t}+z_{i}'f_t+\varepsilon_{i,t}$. Tests and estimation done with $S_{i,t}$ and $ O_{i,t}$  added as cross-section averages and heteroskedastic and autocorrelation robust standard errors.}
    \label{table:fh_breaks}
\end{table}

\begin{table}[]
 \centering
    \begin{tabular}{c ccc}\hline\hline
   \multirow{2}{*}{} & \multirow{2}{*}{Savings} & \multirow{2}{*}{Openness} & Savings $\times$ \\ & & & Openness \\ \hline
        1980 - 1996 & 0.425*** & 0.088*** & -0.002** \\
& (0.116) & (0.022) & (0.001) \\
1997 - 2019 & 0.343*** & 0.098*** & -0.002** \\
& (0.092) & (0.028) & (0.001) \\
\hline\hline

    \end{tabular}
    \caption{Model in \eqref{app:mod} with one structural break in $1996$, estimated by CCEP.}
    \label{table:fh_breaks_estim}
\end{table}

\clearpage

\clearpage
\section{Conclusions}
This paper presents three forms of unobserved heterogeneity, which are significantly more general than the two-way error component model, which is currently the workhorse in the literature. IFE and NSTW nest the TWFE model as a special case and thus applying IFE and NSTW estimators by default can lead to inefficiency but not inconsistency. However, applying TWFE in the presence of IFE, GFE, or NSTW leads to inconsistency. 

The main costs to pay for the new methodologies are inefficiency and computational time. These two costs, however, don't work in the same direction; rather than colluding, they are competing. Large datasets remove the fear of inefficiency, which can lead to low power or wide confidence intervals, but increase computational time. On the contrary, small datasets need careful modeling to gain efficiency through parsimony, but then the computational burden is small.

As such, the main conclusion we reach is that the new methods should be used in applied research. This conclusion is corroborated by the empirical findings in the two applications, which suggest that the unobserved heterogeneity takes more general forms than TWFE and that estimators based on the new models are more appropriate. The two applications studied earlier demonstrate that there is a host of available software already. A comprehensive list can be found in the appendix below.

\bibliographystyle{ecta}
\bibliography{bib.bib}  

\clearpage
\appendix
\section{List of Available Software}
\label{section::app::software}
\setcounter{table}{0}
\begin{table}[!h]\resizebox{\textwidth}{!}{%
    \centering
    \begin{tabular}{l l l}\hline\hline
         & Stata &  R \\ \hline
         \multicolumn{3}{c}{\textit{Estimators}} \\ \hline
         \textbf{(D)CCE}& \textsc{xtdcce2}$^\dagger$, \textsc{xttacce} & \textsc{plm} \\
         \citep{Pesaran2006,ChudikPesaranJoE2015,westerlundkadkara2025} & \citep{Ditzen2019a,Ditzen2021,Chen2025XTTACCE} & \citep{plmR} \\
         \textbf{ILS} &  \textsc{regife} & \textsc{interFE} \\
         \citep{Bai2009,moonweidner2017} & \citep{regife} & \citep{gsynth} \\
         \textbf{TSIV} &\textsc{xtivdfreg}  &  \\
         \citep{NorkuteEtal2021}& \citep{Kripfganz2021} & \\
         \textbf{TSGF-KT}  & $^\bigstar$ & \textsc{pcluster} \\
         \citep{Beyhum2024}& & \citep{pcluster} \\
         \textbf{GF} & $^\spadesuit$ &  \\
         \citep{Bonhomme2015a} &  &  \\  \hline
         \multicolumn{3}{c}{\textit{Test for Cross-Section Dependence and Exponent of CSD}} \\ \hline
         \textbf{CD} &  & \textsc{plm} \\
         \citep{PesaranCD2015,Pesaran2021} &  \textsc{xtcd2}& \citep{plmR} \\
         \textbf{CD*, CDw} & \citep{Ditzen2021}&  \\
         \citep{Pesaran2021b,JuodisReese2021} &  & \\
         \textbf{Exponent,} $\hat{\alpha}$ & \textsc{xtcse2}  &  \textsc{plm}$^\clubsuit$ \\
         \citep{BaileyKapetaniosPesaran2016,BaileyKapetaniosPesaran2019} &\citep{Ditzen2021} & \citep{plmR} \\ \hline
          \multicolumn{3}{c}{\textit{Estimation of Number of Factors}} \\ \hline
         \textbf{PC(m) / IC(m)} & & \textsc{getnfac} \\
         \citep{Bai2002} & & \citep{PANICr}\\
         $\mathbf{\tilde{k}}_{ER}$, $\mathbf{\tilde{k}}_{GR}$ & & \textsc{pc\_fn}\\
         \citep{AhnHorenstein2013} & \textsc{xtnumfac}& \citep{HDRFA}\\
         \textbf{ED} & \citep{ReeseDitzen2023} & \\
         \citep{Onatski2010} & & \\
         $\mathbf{\hat{k}}_{GOS}$ & & $^\sharp $\\
         \citep{Gagliardini2019} & &  \\
         \hline
          \multicolumn{3}{c}{\textit{Tests for Slope Heterogeneity}} \\ \hline
          \textbf{Cross-Section} & \textsc{xthst} &  \textsc{pvar} \\
          \citep{PesaranYamagata2008,Blomquist2013} & \citep{BersvendsenDitzen2021} & \citep{plmR}\\
          \textbf{Time} &\textsc{xtbreak} & \\
          \citep{NKW2022,DKW2025} & \citep{DKWxtbreak2025} & \\
           \textbf{Thresholds} & \textsc{xtthreshold} & \\
           \citep{BarassiKaraviasZhu2025,DKW2025threshold} &  \citep{DKW2025xtthres} & \\
         \hline\hline
    \end{tabular}}
    \caption{Software in Stata and R. Notes:\\$^\dagger$ Includes regularized CCEP \citep{Juodis2022}, test for rank condition \citep{devossarafidis2024}, IC to select cross-section averages \citep{LucaMargaritella2023}. \\ $^\bigstar$ On request. \\ $^\spadesuit$  Code is available on Stephane Bonhomme's webpage, Matlab code on request. \\ $^\clubsuit$  To be released soon. \\  $^\sharp $ Matlab code available on Oliver Scaillet's webpage.}
    \label{tab:software}
\end{table}

\end{document}